\documentclass[twocolumn]{autart}
\usepackage{amsmath}
\usepackage{graphicx}
\usepackage{amsfonts}
\usepackage{amssymb}
\usepackage{graphicx}
\usepackage{mathptmx}
\usepackage{caption}
\usepackage{subfigure}
\usepackage{color}
\usepackage{url}
\usepackage{cite}
\usepackage{CJK}
\usepackage{times}
\newcommand{\T}{^{\mbox{\tiny T}}}

\newtheorem{assumption}{Assumption}

\newtheorem{lemma}{Lemma}

\newtheorem{remark}{Remark}

\newenvironment{proof}[1][\bf Proof]%
{\smallskip\par\noindent\textit{#1: }}%
{\hspace*{\fill} \rule{6pt}{6pt}\smallskip}

\def\diag{{\rm diag}}

\def\th{\mbox{\small th}}
\def\T{^{\rm\tiny T}}

\allowdisplaybreaks

\begin{document}
\begin{frontmatter}

\title{Coordinated Output Regulation of Heterogeneous Linear Systems under Switching Topologies\thanksref{footnoteinfo}}

\thanks[footnoteinfo]{This paper was not presented at any IFAC
meeting. A preliminary version was presented at the 52nd Control and Decision Conference \cite{MengZiyang_CDC13}. Corresponding author Z. Meng. Tel. +46-722-839377.}

\author[KTH]{Ziyang Meng}\ead{ziyangm@kth.se},
\author[KTH]{Tao Yang}\ead{taoyang@kth.se},
\author[KTH]{Dimos V. Dimarogonas}\ead{dimos@kth.se},
\author[KTH]{Karl H. Johansson}\ead{kallej@kth.se}

\address[KTH]{ACCESS Linnaeus Centre, School of Electrical Engineering,
Royal Institute of Technology, Stockholm 10044, Sweden.}

\begin{keyword}
Heterogeneous linear dynamic systems; Coordinated output regulation; Switching communication topology
\end{keyword}

\begin{abstract}
This paper constructs a framework to describe and study the
coordinated output regulation problem for multiple heterogeneous
linear systems. Each agent is modeled as a general linear multiple-input
multiple-output system with an autonomous exosystem which
represents the individual offset from the
group reference for the agent. The multi-agent system as a whole
has a group exogenous state which represents the tracking reference for the whole group.
Under the constraints that the group exogenous output is only locally available to
each agent and that the agents have only access to their neighbors' information,
we propose observer-based feedback controllers to solve the coordinated output regulation
problem using output feedback information. A high-gain approach is used and
the information interactions are allowed to be switched over a finite set of fixed
networks containing both graphs that have a directed spanning tree and graphs that do not.
The fundamental relationship between the information interactions, the dwell time,
the non-identical dynamics of different agents, and the high-gain parameters is given.
Simulations are shown to validate the theoretical results.
\end{abstract}
\end{frontmatter}

\section{Introduction} \label{sec-intro}

Coordinated control of multi-agent systems has recently drawn large attention due to its broad applications in physical, biological, social, and mechanical systems \cite{Cortes_TAC2006,TannerJadbabaiePappas07,Chopra_TAC09,BaiHe_Book}.
The key idea of ``coordination'' algorithm is to realize a global emergence using only local information interactions \cite{JadbabaieLinMorse03,SaberFaxMurray07_IEEE}. The coordination problem of a single-integrator network is fully studied with an emphasis on the system robustness to the input time delays and switching communication topologies \cite{JadbabaieLinMorse03,SaberFaxMurray07_IEEE,Blondel_CDC2005,RenBeard05_TAC}, discrete-time dynamical models \cite{Moreau_TAC05,YouKeyou_TAC2011}, nonlinear couplings \cite{LinZhiyun_SIAM07}, the convergence speed evaluation \cite{CaoMing_SIAM08}, the effects of quantization \cite{CaiKai_TAC11}, and the leader-follower tracking \cite{ShiGuodong_TAC12}.

Following these ideas, the study of coordination of multiple linear dynamic systems becomes an attractive and fruitful research direction for the control community recently. For example, the authors of \cite{PeterWieland_IJSS} generalize the existing works on coordination of multiple single-integrator systems to the case of multiple
linear time-invariant single-input systems.
For a network of neutrally stable systems and polynomially unstable systems, the author of \cite{EmreTuna_TAC09} proposes a design scheme for achieving synchronization.
The case of switching communication topologies is considered in \cite{LucaScardovi_Automatica09} and a so-called consensus-based
observer is proposed to guarantee leaderless synchronization of multiple identical linear dynamic systems under a jointly connected communication topology. Similar problems are also considered
in \cite{ChenDaizhan_AJC2008} and \cite{ChengDaizhan_SCL2010}, where a frequently connected communication topology is studied in \cite{ChenDaizhan_AJC2008} and an assumption on the neutral stability is imposed in \cite{ChengDaizhan_SCL2010}.
The authors of \cite{Lizhongkui_ITCSI10} propose a neighbor-based observer to solve the synchronization problem for general linear time-invariant systems.
An individual-based observer and a low-gain technique are used in
\cite{HyungboShim_Automatica09} to synchronize a group of linear systems with open-loop poles at most polynomially unstable.
In addition, the classical Laplacian matrix is generalized in
\cite{YangTao_IJRNC2011} to a so-called interaction matrix. A D-scaling approach is then used to stabilize this interaction matrix under both fixed and switching communication topologies.
Synchronization of multiple heterogeneous linear systems has been investigated under both fixed and switching communication topologies \cite{Lunze_TAC12,ChenXiang_CDC13,Wieland_Automatica2011}. A similar problem is studied in \cite{Havard_Automatica2012,YangTao_IJRNC2013}, where a high-gain approach is proposed to dominate the non-identical dynamics of the agents.
The cases of frequently connected and jointly connected communication topologies are studied in \cite{Hongkeun_CDC2010} and \cite{Hongkeun_Automatica2013}, respectively, where a slow switching condition and a fast switching condition are presented.
Recently, the generalizations of coordination of multiple linear dynamic systems to the cooperative output
regulation problem are studied in \cite{WangXiaoli_TAC2010,SuYoufeng_TAC2012a,DingZhengtao_TAC13}. In addition, the study on the synchronization of homogenous or heterogeneous networks with nonlinear couplings also attracts extensive attention \cite{CaoJinde_PhysicaA2007,CaoJinde_SMCB2008,CaoJinde_PhysicaA2010,HeWangli_Neuro2013}.

In this paper, we generalize the classical output regulation
problem of an individual linear dynamic system to the coordinated output regulation problem of multiple heterogeneous linear dynamic systems.
We consider the case where each agent has an individual offset and
simultaneously there is a group tracking reference. The individual offset and the group
reference are generated by autonomous systems ({\em i.e.,} systems without inputs).
Each individual offset is available to its corresponding agent while the group reference
can be obtained only through constrained communication among the agents, {\em i.e.,} the
group reference trajectory is available to only a subset of the agents.
Our goal is to find an observer-based feedback controller for each agent such that the output of each agent converges to a given trajectory determined by the combination of the individual offset and the group reference.
Motivated by the approach proposed in \cite{Havard_Automatica2012}, we propose a unified observer to solve the coordinated output regulation problem of multiple heterogeneous general linear dynamics, where the open-loop poles of the agents can be exponentially unstable and the dynamics are allowed to be different both with respect to dimensions and parameters. This
relaxes the common assumption of identical dynamics \cite{EmreTuna_TAC09,LucaScardovi_Automatica09,ChengDaizhan_SCL2010,Lizhongkui_ITCSI10,Hongkeun_CDC2010}
or open-loop poles at most polynomially unstable \cite{LucaScardovi_Automatica09,ChengDaizhan_SCL2010,Wieland_Automatica2011}. The main contribution of this work is that
the information interaction is allowed to be switching from a graph set containing both a directed spanning tree set and a disconnected graph set for the case of heterogeneous linear systems. This extends
the existing works on the case of fixed communication topologies \cite{EmreTuna_TAC09,Lizhongkui_ITCSI10,Havard_Automatica2012,WangXiaoli_TAC2010}.
The high-gain technique is used and
the relationships between the dwell time \cite{Liberzon_ICSM1999},
the non-identical dynamics among different agents and the high-gain parameters are also given.

The remainder of the paper is organized as follows. In
Section~\ref{sec:graph}, we give some basic definitions on network model. In Section~\ref{sec:problem}, we formulate the problem of coordinated output regulation of multiple heterogenous linear systems. We then propose the state feedback control law with a unified observer design
in Section \ref{sec:unified}. Two case studies are given in Section \ref{sec:case-study}.
Numerical
studies are carried out in Section~\ref{sec:simulation} to validate our designs of observer-based controllers
and a brief concluding remark is drawn in
Section~\ref{sec:conclusion}.

\section{Network Model} \label{sec:graph}

We use graph theory to model the communication topology among
agents.
A directed graph ${G}$ consists of
a pair $( \mathbf{{V}}, \mathbf{{E}})$, where $
\mathbf{{V}}=\{\nu_1,\nu_2,\ldots, \nu_{n}\}$ is a finite, nonempty set of nodes and
$ \mathbf{{E} }\subseteq  \mathbf{{V}}\times  \mathbf{{V}}$ is a set of
ordered pairs of nodes. An edge $(\nu_i,\nu_j)$ denotes that node $\nu_j$
can obtain information from node $\nu_i$.
All neighbors of node $\nu_i$ are denoted as $N_i :=
\{\nu_j|(\nu_j,\nu_i)\in  \mathbf{{E}}\}$.
For an edge $(\nu_i,\nu_j)$ in a directed graph, $\nu_i$ is the parent node and $\nu_j$ is the child node.
A directed path in a directed graph is a sequence of edges of the form $(\nu_{i},
\nu_{j}),(\nu_{j}, \nu_{k}),\ldots$. A directed tree
is a directed graph, where every node has exactly one parent except
for one node, called the root, which has no parent, and the root has
a directed path to every other node.
A directed graph has a directed spanning tree if there
exists at least one node having a directed path to all
other nodes.

For a leader-follower graph
$\overline{{G}}:= (\overline{ {V}},\overline{ {E}})$, we have $\overline{
{V}}=\{\nu_0,\nu_1,\dots, \nu_{n}\}$,
$\overline{ {E}} \subseteq \overline{ {V}}\times \overline{ {V}}$, where $\nu_0$ is the leader and $\nu_1,\nu_2,\ldots, \nu_{n}$ denote the followers.
The leader-follower adjacency matrix
$\overline{{A}}=[a_{ij}]\in \mathbb{R}^{(n+1) \times (n+1)}$ is defined such that $a_{ij}$ is positive if $(\nu_j,\nu_i) \in
\overline{ {E}}$ while $a_{ij}=0$ otherwise. Here we assume that $a_{ii}=0$, $
i=0,1,\dots,n$, and the leader has no parent, {\em i.e.}, $a_{0j}=0,j=0,1,\cdots,n$. The leader-follower ``grounded'' Laplacian matrix $L = [l_{ij}] \in \mathbb{R}^{{n \times n}}$ associated with
$\overline{ {A}}$ is defined as $l_{ii}=\sum_{j=0}^na_{ij}$ and
$l_{ij}=-a_{ij}$, where $i\neq j$.

In this paper, we assume that the leader-follower communication topology $\overline{{G}}_{\sigma(t)}$ is time-varying and switching from a finite set $\{\overline{{G}}_k\}_{k\in \Gamma}$, where $\Gamma=\{1,2,\dots,\delta\}$ is an index set and $\delta\in \mathbb {N}$ indicates its cardinality. We impose the technical condition that $\overline{{G}}_{\sigma(t)}$ is right continuous, where ${\sigma}:[t_0,\infty)\rightarrow\Gamma$ is a piecewise constant function of time. That is to say, $\overline{{G}}_{\sigma(t)}$ remains constant for
$t\in[t_{\ell},t_{\ell+1})$, $\ell=0,1,\dots$ and switches at $t=t_{\ell}$,
$\ell=1,2,\dots$. In addition, we assume that $\inf_{\ell}(t_{\ell+1}-t_\ell)\geq \tau_d>0$, $\ell=0,1,\dots$, with $\lim_{\ell\rightarrow\infty}t_{\ell}=\infty$, where $\tau_d$ is a constant known as the dwell time \cite{Liberzon_ICSM1999}.

Let the sets $\{\overline{{A}}_k\}_{k\in \Gamma}$ and $\{L_k\}_{k\in \Gamma}$ be the leader-follower adjacency matrices and leader-follower grounded Laplacian matrices associated with $\{\overline{{G}}_k\}_{k\in \Gamma}$, respectively.
Consequently, the time-varying leader-follower adjacency matrix and time-varying leader-follower grounded Laplacian matrix are defined as $\overline{{A}}_{\sigma(t)}= [a_{ij}(t)]$ and $L_{\sigma(t)}= [l_{ij}(t)]$.

Other notation in this paper:
$\lambda_{\min}(P)$ and $\lambda_{\max}(P)$ denote, respectively,
the minimum and maximum eigenvalues of a real symmetric matrix $P$, $P\T$ denotes the transpose of $P$, and $I_{n}$ denotes the $n\times n$ identity matrix.

\section{Problem Formulation}\label{sec:problem}

\subsection{Agent Dynamics}

Suppose that we have $n$ agents modeled by the linear MIMO systems:
\begin{equation}
\dot x_i=A_ix_i+B_iu_i, \label{eq:state}
\end{equation}
where $x_i \in \mathbb{R}^{n_i}$ is the agent state,
$u_i \in \mathbb{R}^{m_i}$ is the control input,
$A_i\in \mathbb{R}^{n_i\times n_i}$, and
$B_i\in \mathbb{R}^{n_i\times m_i}$.

Also suppose that there is an individual autonomous exosystem for each $\nu_i\in \mathbf{V}$,
\begin{equation}
\dot \omega_i=S_i\omega_i, \label{eq:disturbance}
\end{equation}
where $\omega_i\in \mathbb{R}^{q_i}$ and $S_i\in \mathbb{R}^{q_i\times q_i}$.

In addition, there is a group autonomous exosystem for the multi-agent system as a whole:
\begin{equation}
\dot x_0=A_0x_0, \label{eq:leader}
\end{equation}
where $x_0\in \mathbb{R}^{n_0}$ and $A_0\in \mathbb{R}^{n_0\times n_0}$.

\subsection{Control Architecture}\label{sec:architecture}

\begin{figure}
\begin{center}
\includegraphics[scale=0.45]{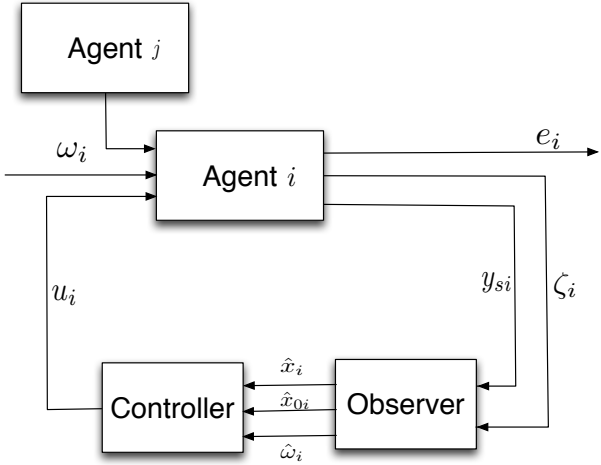}
  \caption{Control architecture for agent $\nu_i$}
\label{fig:architecture}
 \end{center}
 \end{figure}
The control of each agent is supposed to have the structure shown in Fig. \ref{fig:architecture}.
More specifically, for the individual autonomous exosystem tracking, available output information for agent $\nu_i\in \mathbf{V}$ is
\begin{equation*}
y_{si}=C_{si}x_i+C_{wi}\omega_i,
\end{equation*}
where $C_{si}\in \mathbb{R}^{p_1\times n_i}$, and $C_{wi}\in \mathbb{R}^{p_1\times q_i}$.

For the group autonomous exosystem tracking, only neighbor-based output information is available due to the constrained communication. This means that not all the agents have access to $y_0$. The available information is the neighbor-based sum of each agent's own output relative to that of its' neighbors, {\em i.e.,}
\begin{equation*}
\zeta_i=\sum_{j=0}^n a_{ij}(t)(y_{di}-y_{dj})
  \end{equation*}
is available for each agent $\nu_i\in \mathbf{V}$, where $a_{ij}(t)$,
$i=0,1,\ldots,n$, $j=0,1,\ldots,n$, is
entry $(i,j)$ of the adjacency matrix $\overline{{A}}_{\sigma(t)}$ associated with $\overline{{G}}_{\sigma(t)}$ defined
in Section \ref{sec:graph} at time $t$, $y_{di}$ can be represented by
$y_{di}=C_{di}x_i$, $i=1,2,\dots,n$ and $y_{d0}=C_0x_0$,
where $C_{di}\in \mathbb{R}^{p_2\times n_i}$, $i=1,2,\dots,n$ and
$C_0\in \mathbb{R}^{p_2\times n_0}$.
Also, the relative estimation information is available using the same communication topologies, {\em i.e.,}
\begin{equation*}
\widehat{\zeta}_i=\sum_{j=0}^n a_{ij}(t)(\widehat{y}_{i}-\widehat{y}_{j})
  \end{equation*}
is available for each agent $\nu_i\in \mathbf{V}$, where $\widehat{y}_{i}$ is an estimation produced internally by each agent $\nu_i\in \mathbf{V}$.

Fig. \ref{fig:COMMTOP1} gives an example of information flow among the agents and the group autonomous exosystem $\nu_0$ for $n=3$ agents.

\subsection{Switching Topologies}
For the communication topology set $\{\overline{{G}}_k\}_{k\in \Gamma}$, we assume that $\overline{{G}}_k$, $\forall k\in\Gamma_c$ is a graph containing a directed spanning tree with $\nu_0$ rooted. Without loss of generality, we relabel $\Gamma_c:=\{1,2,\dots,\delta_1\}$ ($1\leq \delta_1\leq\delta$), where $\delta_1\in\mathbb {N}$. The remaining graphs are labeled as $\overline{{G}}_k$, $\forall k\in\Gamma_d$, where $\Gamma_d:=\{\delta_1+1,\delta_1+2,\dots,\delta\}$. Denote the graph set $\overline{\mathbb{G}}_c=\{\overline{{G}}_k\}_{k\in \Gamma_c}$ and the graph set
$\overline{\mathbb{G}}_d=\{\overline{{G}}_k\}_{k\in \Gamma_d}$, respectively.
We also denote $T^d_{\overline{t}_0}(t)$ and $T_{\overline{t}_0}^c(t)$ the total activation time when $\overline{{G}}_{\sigma(\varsigma)}\in\overline{\mathbb{G}}_d $ and total activation time when $\overline{{G}}_{\sigma(\varsigma)}\in\overline{\mathbb{G}}_c$ during $\varsigma\in[\overline{t}_0,t)$ for $\overline{t}_0\geq t_0$.

\begin{assumption}\label{assm:output-dwell}
The dwell time $\tau_d$ is a positive constant.
\end{assumption}

\begin{assumption}\label{assm:output-time}
Given a positive constant $\kappa$, there exists a
$\overline{t}_0\geq t_0$ such that $T^c_{\overline{t}_0}(t)\geq \kappa T^d_{\overline{t}_0}(t)$ for all $t\geq \overline{t}_0$.
\end{assumption}
\begin{remark}
Note that a sufficient condition satisfying
Assumption \ref{assm:output-time} is that $\overline{\mathbb{G}}_c$
is non-empty and given a $T>0$ and $\tau_d>0$, for any $t\geq t_0$,
the switching signal $\sigma(t)$ satisfies
$\{t| \overline{{G}}_{\sigma(t)}\in \overline{\mathbb{G}}_c\} \cap [t,t+T]\neq \emptyset$.
Such a condition is also referred as ``frequently connected'' condition
({\em i.e.,} the communication topology
that contains a directed spanning tree is active frequently
enough \cite{ChenDaizhan_AJC2008,YangTao_IJRNC2011}). Note that this condition
implies that there exists a time sequence $0=T_0<T_1<\dots<T_\ell\dots$
such that $\{t| \overline{{G}}_{\sigma(t)}\in \overline{\mathbb{G}}_c\}
\cap [T_\ell,T_{\ell+1}]\neq \emptyset$, for all $\ell=0,1,\dots$,
where $T_{\ell+1}-T_\ell\leq 2T$. Therefore, there exists a $\overline{t}_0\in[t_0,t_0+2T]$
such that $T^c_{\overline{t}_0}(t)\geq \frac{\tau_d}{2T} T^d_{\overline{t}_0}(t)$ for all $t\geq \overline{t}_0$.
\end{remark}

\subsection{Control Objective}\label{sec:objective}

The control objective of each agent is to track a given trajectory determined by the combination of the group reference
$x_0$ and the individual offset $\omega_i$, $i=1,2,\dots,n$. Such a combination is captured by the coordinated output regulation tracking error ({\em i.e.,} the total tracking error representing the combination of both individual tracking and group tracking of each agent):
\begin{equation}
e_i=D_{si}x_i+D_{wi}\omega_i+D_{0}x_0.\label{eq:output-error}
  \end{equation}
Thus, our objective is to guarantee that $\lim_{t\rightarrow\infty} e_{i}(t)=0$.
We design an observer-based controller with available individual output information and neighbor-based group output information to solve this problem.

\begin{figure}
\begin{center}
\includegraphics[scale=0.4]{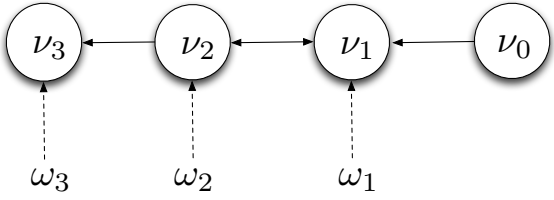}
\caption{Information flow associated with three agents $\nu_1$, $\nu_2$, $\nu_3$, the individual autonomous exosystems
$\omega_1$, $\omega_2$, $\omega_3$, and the group autonomous exosystem $\nu_0$}
\label{fig:COMMTOP1}
 \end{center}
 \end{figure}

For the system shown in Fig. \ref{fig:COMMTOP1}, the overall control can correspond to a formation control problem,
where $\omega_i$ encodes the relative position between each agent and the leader while the leader $x_0$ defines the overall motion of the group.

\section{Coordinated Output Regulation with Unified Observer Design}\label{sec:unified}

As suggested by Fig. \ref{fig:architecture}, the design procedure to
solve the coordinated output regulation problem includes two main steps:
the first one is the state feedback control design and
the second one is the
observer design for the group autonomous exosystem, the individual autonomous exosystem,
and internal state information for each agent.

\subsection{Redundant Modes}\label{sec:remove1}

Before designing state feedback
control and distributed observer, we need first to remove the redundant modes that
have no effect on $y_{si}$ and $y_{di}-y_{d0}$.

We impose the following assumptions on the structure of the systems.
\begin{assumption}\label{assm:A-C}
$\newline$
$\bullet$ $\left(A_i,\left[\begin{array}{c} C_{si} \\C_{di}\end{array} \right]\right)$, $i=1,2,\dots,n$ is observable.
\newline
$\bullet$ $(S_i,C_{wi})$, $i=1,2,\dots,n$ is observable.
\newline
$\bullet$ $(A_0,C_0)$, $i=1,2,\dots,n$ is observable.
\end{assumption}

We first write the state and output of each agent in the compact form
\begin{subequations}
\begin{align*}
\left[\begin{array}{c} \dot x_i\\ \dot \omega_i \\ \dot x_0\end{array} \right]
&=\left[\begin{array}{ccc} A_i & 0& 0\\ 0& S_i & 0 \\0& 0&A_0\end{array} \right]
\left[\begin{array}{c} x_i\\ \omega_i \\ x_0\end{array} \right]
+\left[\begin{array}{c} B_i\\ 0 \\ 0\end{array} \right]u_i,
\\
\left[\begin{array}{c} y_{si}\\ y_{di}-y_{d0}\end{array} \right]&=\left[\begin{array}{ccc} C_{si} & C_{wi} &0\\ C_{di}& 0 & -C_0 \end{array} \right]
\left[\begin{array}{c} x_i\\ \omega_i \\ x_0\end{array} \right].
\end{align*}
\end{subequations}
Given that Assumption \ref{assm:A-C} is satisfied, we can perform the state transformation given in Step 1 of \cite{Havard_Automatica2012} by considering $\omega_i$ and $x_0$ together.
We can construct a new state $\overline{x}_i=W_i
\left[\begin{array}{c} x_i\\ \omega_i \\ x_0\end{array} \right]$ with the dynamics
\begin{subequations}\label{eq:new-trans-closed}
\begin{align}
\dot {\overline{x}}_i&=\overline{A}_i\overline{x}_i+\overline{B}_iu_i=
\left[\begin{array}{cc} A_i & \overline{A}_{i12}\\ 0& \overline{A}_{i22} \end{array} \right]\overline{x}_i
+\left[\begin{array}{c} B_i\\ 0\end{array} \right]u_i,
\\
\left[\begin{array}{c} y_{si}\\ e_{di}\end{array} \right]&=\overline{C}_i\overline{x}_i
=\left[\begin{array}{cc} C_{si} & \overline{C}_{i21}\\ C_{di} & \overline{C}_{i22} \end{array} \right]\overline{x}_i.
\end{align}
\end{subequations}
where $e_{di}=y_{di}-y_{d0}$, and the details designs on $W_i$, $\overline{A}_i$, $\overline{B}_i$, $\overline{C}_i$
are given in \cite{Havard_Automatica2012}.
It was shown that pair $(\overline{A}_i,\overline{C}_i)$ is observable
and the eigenvalues of $\overline{A}_{i22}$ are a subset of the eigenvalues of $S_i$ and $A_0$, $i=1,2,\dots,n$.

\subsection{Regulated State feedback Control Law}\label{sec:controller}

We now design a controller to regulate $e_i$ to zero for each agent based on the state
information $\overline{x}_i=\left[
\begin{array}{c} \overline{x}_{i1}\\ \overline{x}_{i2}\end{array} \right]$,
where $\overline{x}_{i1}\in \mathbb{R}^{n_i}$.

We impose the following assumptions on the structure of the systems.
\begin{assumption}\label{assm:A-B}
$\newline$
$\bullet$ $(A_i,B_i)$ is stabilizable, $i=1,\dots,n$.
$\newline$
$\bullet$ $(A_i,B_i,D_{si})$ is right-invertible, $i=1,\dots,n$.
$\newline$
$\bullet$ $(A_i,B_i,D_{si})$ has no invariant zeros in
the closed right-half complex plane that coincide with the eigenvalues of $S_i$ or $A_0$, $i=1,\dots,n$.
\end{assumption}

\begin{lemma}\label{lem:control-full}
Let Assumption \ref{assm:A-B} hold. Then,
the regulator equations \eqref{eq:output-regulator} are solvable and
the state-feedback controller
$u_i=F_i(\overline{x}_{i1}-\Pi_{i}\overline{x}_{i2})+\Gamma_{i}\overline{x}_{i2}$
ensures that $\lim_{t\rightarrow\infty}e_{i}(t)=0$, $i=1,2\ldots,n$,
where $\Pi_{i}$, $\Gamma_{i}$ are the solutions of the following regulator equations
\begin{subequations}\label{eq:output-regulator}
  \begin{align}
\Pi_{i}\overline{A}_{i22}&=A_i\Pi_{i}+\overline{A}_{i12}+B_i\Gamma_{i}, \\
0&=D_{si}\Pi_{i}+\left[
\begin{array}{cc} D_{wi}&D_0 \end{array} \right], \quad i=1,2\dots,n,
  \end{align}
\end{subequations}
and $F_i$ is chosen such that $A_i+B_iF_i$ is Hurwitz.
\end{lemma}
\proof
It follows from \cite{Saberi_book2003} and the similar analysis of proof of Lemma 3 in \cite{Havard_Automatica2012},
we can show that the regulator equations \eqref{eq:output-regulator}
are solvable given that Assumption \ref{assm:A-B} is satisfied. Then, by considering
$\dot {\overline{x}}_{i2}=\overline{A}_{i22}\overline{x}_{i2}$ as
the exosystem and $\dot x_i=A_ix_i+B_iu_i$ as the system to be regulated for the
classic output regulation result \cite{Francis_SIAM77}, we know that
$u_i=F_i(\overline{x}_{i1}-\Pi_{i}\overline{x}_{i2})+\Gamma_{i}\overline{x}_{i2}$
ensures that $\lim_{t\rightarrow\infty}e_i(t)=0$, $i=1,2\ldots,n$,
where $\Pi_{i}$ and $\Gamma_{i}$ are the solutions of the regulator
equations \eqref{eq:output-regulator}.
\endproof

We next design observers to estimate $\overline{x}_i$
based on output information $y_{si}$ and $\zeta_i$ for each agent.

\subsection{Pseudo-identical Linear Transformation}

Note that the individual offset $\omega_i$ can be estimated by $y_{si}$ and
the group reference $x_0$ can be estimated by $\widehat{\zeta}_i$. In contrast,
the internal state information $x_i$ for each agent can be obtained by either
$y_{si}$ or $\widehat{\zeta}_i$.
In this section, we use the combination of $y_{si}$ and $\widehat{\zeta}_i$
to give a unified observer design.

We define $\chi_i=T_i\overline{x}_i\in\mathbb{R}^{p\overline{n}}$, $i=1,2,\dots,n$,
where $\overline{n}=n_0+\max_{i=1,2,\dots,n}(n_i+q_i)$, $p=p_1+p_2$, and
\begin{equation*}
T_i=\left[
\begin{array}{c} \overline{C}_i \\ \vdots \\
\overline{C}_i\overline{A}_i^{\overline{n}-1} \end{array} \right].
\end{equation*}
Note that $T_i$ is full column rank since the pair $(\overline{A}_i,\overline{C}_i)$,
$i=1,2,\dots,n$ is observable. This implies that $T_i\T T_i$ is nonsingular.
Therefore, it follows that
\begin{subequations}\label{eq:new-trans-close}
  \begin{align}
\dot\chi_i&=(\mathcal{A}+\mathcal{L}_i)\chi_i+\mathcal{B}_iu_i,
\\
\left[\begin{array}{c} y_{si}\\ e_{di}\end{array} \right]&=\mathcal{C}\chi_i,\quad i=1,2,\dots,n,
  \end{align}
\end{subequations}
where $\mathcal{A}=\left[
\begin{array}{cc} 0 & I_{p(\overline{n}-1)}\\ 0 & 0 \end{array} \right]\in \mathbb{R}^{p\overline{n}\times p\overline{n}}$,
$\mathcal{L}_i=\left[
\begin{array}{c} 0 \\  L_i \end{array} \right]$, 
$\mathcal{B}_i=T_i\overline{B}_i$, $\mathcal{C}=\left[
\begin{array}{cc} I_p & 0 \end{array} \right]\in \mathbb{R}^{p\times p\overline{n}}$ for some matrix $L_i\in\mathbb{R}^{p\times p\overline{n}}$.

\subsection{Unified Observer Design}

Motivated by \cite{Havard_Automatica2012}, based on the available output information
$y_{si}$ and the neighbor-based group
output information $\zeta_i$,
the distributed observer is proposed for \eqref{eq:new-trans-close} as
\begin{align}
&\dot{\widehat{\chi}}_i=(\mathcal{A}+\mathcal{L}_i)\widehat{\chi}_i+\mathcal{B}_iu_i
+S(\varepsilon)\mathcal{P}\mathcal{C}\T
\notag\\ &\times\left(\left[
\begin{array}{c} y_{si}\\ \sum_{j=0}^na_{ij}(t)(y_{di}-y_{dj}) \end{array} \right]
-\left[
\begin{array}{c} \widehat{y}_{si}\\
\sum_{j=0}^na_{ij}(t)(\widehat{y}_i-\widehat{y}_j)\end{array} \right]
\!\!\right),
\notag\\ &\qquad\qquad\qquad\qquad\qquad\qquad i=1,2\ldots,n, \label{eq:observer-full}
\end{align}
where $a_{ij}(t)$, $i=0,1,\ldots,n$, $j=0,1,\ldots,n$, is
entry $(i,j)$ of the adjacency matrix $\overline{A}_{\sigma(t)}$ associated with
$\overline{{G}}_{\sigma(t)}$ defined
in Section \ref{sec:graph} at time $t$, $\widehat{y}_{si}=\mathcal{C}_1\widehat{\chi}_i$,
$\widehat{y}_i=\mathcal{C}_2\widehat{\chi}_i$, $i=1,\ldots,n$, $\mathcal{C}_1$ is first $p_1$ rows of
$\mathcal{C}$, $\mathcal{C}_2$ is the remaining $p_2$ rows of
$\mathcal{C}$, and $\widehat{y}_0=0$.
In addition, $S(\varepsilon)=\diag(I_p\varepsilon^{-1},I_p\varepsilon^{-2},\dots,I_p\varepsilon^{-\overline{n}} )$, where $\varepsilon\in (0,1]$ is a positive constant to be determined, and $\mathcal{P}=\mathcal{P}\T$ is a positive definite matrix satisfying
\begin{align}
\mathcal{A}\mathcal{P}+\mathcal{P}\mathcal{A}\T-2 \mathcal{P}\mathcal{C}\T\left[
\begin{array}{cc} I_{p_1}&0\\
0& \theta I_{p_2}\end{array} \right]\mathcal{C} \mathcal{P}+I_{p\overline{n}}=0,
\label{eq:P-full}
\end{align}
where $\theta=\min_{k\in \Gamma_c}\beta_k$ and $\beta_k$ will be determined later.
Note that the existence of $\mathcal{P}$ is due to the fact that
$\left(\mathcal{A},\left[\begin{array}{cc} I_{p_1}&0\\
0& \sqrt{\theta} I_{p_2}\end{array} \right]\mathcal{C}\right)$ is observable.

\begin{lemma}\label{lem:graph}
\begin{itemize}

\item All the eigenvalues of $L_k$ are in the closed right-half
plane with those on the imaginary axis being simple, where $L_k$ is associated with $\overline{{G}}_k$ defined in Section \ref{sec:graph}, and some $\overline{{G}}_k\in \{\overline{{G}}_k\}_{k\in\Gamma}$.
  \item Furthermore, all the eigenvalues of $L_k$ are in the open right-half
plane for $\overline{{G}}_k\in \{\overline{{G}}_k\}_{k\in\Gamma_c}$.
\end{itemize}
\end{lemma}
\proof
See Theorem 4.29 in \cite{QuZhihua_Book} and Lemma 1.6 in \cite{RenCao11_Book}.
\endproof

\begin{lemma}\label{lem:observer-full}
Let Assumptions \ref{assm:output-dwell}, \ref{assm:output-time},
\ref{assm:A-B}, and \ref{assm:A-C} hold and
assume that $\kappa\geq \frac{\alpha+4\max\{\theta,1\}\lambda_{\max}^2(\mathcal{P})}{1-\alpha}$,
where $\alpha\in (0,1)$. Then, there exists an $\varepsilon^*\in (0,1]$
\footnote{The upper bound of the high-gain parameter may be conservative.
We can use an empirical approach to derive a feasible $\varepsilon^*$
in the practical applications.}
such that,
if $\varepsilon\in (0,\varepsilon^*]$,
$\lim_{t\rightarrow\infty}(\chi_i(t)-\widehat\chi_i(t))=0$, $i=1,2\ldots,n$,
for systems \eqref{eq:observer-full}.
\end{lemma}

\proof
Note that for all $i=1,2,\dots,n$, $\sum_{j=0}^na_{ij}(t)(y_{di}-y_{dj})
=\sum_{j=1}^nl_{ij}(t)((y_{dj}-y_{d0})=\sum_{j=1}^nl_{ij}(t)e_{dj}$.
Define $\widetilde{\chi}_i=\chi_i-\widehat{\chi}_i$. It then follows from \eqref{eq:new-trans-close} and \eqref{eq:observer-full} that
\begin{align*}
\dot{\widetilde{\chi}}_i=&(\mathcal{A}+\mathcal{L}_i)\widetilde{\chi}_i-S(\varepsilon)
\mathcal{P}\mathcal{C}\T
\left(\!\left[
\begin{array}{c} y_{si}-\widehat{y}_{si}\\
\sum_{j=1}^nl_{ij}(t)(e_{dj}-\widehat y_j)\end{array} \right]\!\!\right),
\notag\\ &\qquad\qquad\qquad\qquad\qquad\qquad\ i=1,2\ldots,n,
\end{align*}
where $l_{ij}(t)$, $i=1,\ldots,n$, $j=1,\ldots,n$, is
the $(i,j)\th$ entry of the adjacency matrix $L_{\sigma(t)}$ associated with $\overline{{G}}_{\sigma(t)}$ defined
in Section \ref{sec:graph} at time $t$.
It follows that
\begin{align*}
\dot{\widetilde{\chi}}_i=&(\mathcal{A}+\mathcal{L}_i)\widetilde{\chi}_i-S(\varepsilon)
\mathcal{P}\mathcal{C}\T \left(\!\left[
\begin{array}{c} \mathcal{C}_1\widetilde{\chi}_i\\
\mathcal{C}_2\sum_{j=1}^nl_{ij}(t)\widetilde{\chi}_j\end{array} \right]\!\!\right),
\notag\\ &\qquad\qquad\qquad\qquad\qquad\qquad\qquad\qquad i=1,2\ldots,n.
\end{align*}
By introducing $\xi_i=\varepsilon^{-1}S^{-1}(\varepsilon)\widetilde{\chi}_i$ and after some manipulation,
we have that
\begin{align*}
\varepsilon\dot{\xi}_i=(\mathcal{A}+\mathcal{L}_{i\varepsilon})\xi_i
-\mathcal{P}\mathcal{C}\T \left(\!\left[
\begin{array}{c} \mathcal{C}_1\xi_i\\
\mathcal{C}_2\sum_{j=1}^nl_{ij}(t)\xi_j\end{array} \right]\!\!\right),
\notag\\ i=1,2\ldots,n,
\end{align*}
where $\mathcal{L}_{i\varepsilon}=\left[
\begin{array}{c} 0 \\ \varepsilon^{\overline{n}+1}L_iS(\varepsilon) \end{array} \right]
=O(\varepsilon)$.

Note that $\left[
\begin{array}{c} \mathcal{C}_1\xi_i\\
\mathcal{C}_2\xi_i\end{array} \right]=\mathcal{C}\xi_i$, for all $i=1,2,\dots,n$.
The overall dynamics can be written as
\begin{align}
\varepsilon\dot{\xi}=&\left(I_n\otimes \mathcal{A}+\mathcal{L}_{\varepsilon}
-(I_n\otimes\mathcal{P}\mathcal{C}\T)
\right.
\notag\\&\times \left.\left(I_n\otimes\left[
\begin{array}{cc} I_{p_1}&0\\
0& 0\end{array} \right]+L_{\sigma}\otimes\left[
\begin{array}{cc} 0&0\\
0& I_{p_2}\end{array} \right]\right)(I_n\otimes\mathcal{C})
\right)\xi,\label{eq:switch-full}
\end{align}
where $\xi=[\xi_1\T,\xi_2\T,\dots,\xi_n\T]\T$ and $\mathcal{L}_{\varepsilon}=\diag(\mathcal{L}_{1\varepsilon},\mathcal{L}_{2\varepsilon},\dots,\\ \mathcal{L}_{n\varepsilon})$.

Note that $-L_k$, $k\in \Gamma_c$ is a Hurwitz stable matrix according to Lemma \ref{lem:graph}. Therefore, we can always guarantee that $-L_k+\beta_kI_n$ is also a Hurwitz stable matrix by choosing $\beta_k$ sufficiently small. In particular, we choose $\beta_k$ as a positive constant satisfying $\beta_k<\min\Re\{\lambda(L_k)\}$, $k\in \Gamma_c$, where $\min\Re\{\lambda(L_k)\}$ denote the minimum value of all the real parts of the eigenvalues of $L_k$. Then, we define piecewise Lyapunov function candidate $V_k=\varepsilon\xi\T (P_k\otimes \mathcal{P}^{-1})\xi$, where $P_k$ is positive definite matrix satisfying
\begin{subequations}
  \begin{equation*}
P_k(-L_k+\beta_kI_n)+(-L_k+\beta_kI_n)\T P_k=-I_n<0,~ k\in \Gamma_c,
  \end{equation*}
    \begin{equation*}
P_k(-L_k)+(-L_k)\T P_k\leq 0,\qquad k\in \Gamma_d,
  \end{equation*}
\end{subequations}
where the second inequality is due to Lemma \ref{lem:graph}.

It then follows that for all $k\in \Gamma_c$,
\begin{align*}
\dot V_k\leq&~ 2\xi\T\left(P_k\otimes\mathcal{P}^{-1}\mathcal{A}\right)\xi
+2\xi\T\left(P_k\otimes\mathcal{P}^{-1}\right)\mathcal{L}_{\varepsilon}\xi
\\&-2\xi\T\left(P_k\otimes\left(\mathcal{C}\T \left[
\begin{array}{cc} I_{p_1}&0\\
0& 0\end{array} \right] \mathcal{C}\right)\right)\xi
\\&-2\xi\T\left(P_kL_k\otimes\left(\mathcal{C}\T \left[
\begin{array}{cc} 0&0\\
0& I_{p_2}\end{array} \right] \mathcal{C}\right)\right)\xi
\\ \leq & ~\xi\T\left(P_k\otimes\left(\mathcal{P}^{-1}\mathcal{A}+\mathcal{A}\T\mathcal{P}^{-1}
-2\theta \mathcal{C}\T\left[
\begin{array}{cc} 0&0\\
0& I_{p_2}\end{array} \right] \mathcal{C}\right.\right.
\\&\left.\left.-2\mathcal{C}\T\left[
\begin{array}{cc} I_{p_1}&0\\
0& 0\end{array} \right] \mathcal{C} \right)\right)\xi
+2\xi\T\left(P_k\otimes\mathcal{P}^{-1}\right)\mathcal{L}_{\varepsilon}\xi
\\&-\xi\T\left(\left(2P_kL_k-2\theta P_k\right)
\otimes(\mathcal{C}\T\left[
\begin{array}{cc} 0&0\\
0& I_{p_2}\end{array} \right] \mathcal{C})\right)\xi
\\ \leq& ~\xi\T\!\left(P_k\!\otimes\! \left(\mathcal{P}^{-1}
\left(\!\mathcal{A}\mathcal{P}+\mathcal{P}\mathcal{A}\T\right.\right.\right.
\\&\left.\left.\left.-2\mathcal{P}\mathcal{C}\T \left[
\begin{array}{cc} I_{p_1}&0\\
0& \theta I_{p_2}\end{array} \right]\mathcal{C}\mathcal{P}\right)
\mathcal{P}^{-1}\!\right)\!\right)\xi
\\&
-\xi\T\left(\!\left(\!P_kL_k+L_k\T P_k-2\beta_k P_k\!\right)
\otimes(\mathcal{C}\T \left[
\begin{array}{cc} 0&0\\
0& I_{p_2}\end{array} \right]\mathcal{C})\!\right)\xi
\\&+ 2\lambda_{\max}(P_k) \lambda_{\max}(\mathcal{P}^{-1})
\|\mathcal{L}_{\varepsilon}\| \|\xi\|^2
\\ \leq& -\xi\T\left(P_k\otimes (\mathcal{P}^{-1}\mathcal{P}^{-1})\right)\xi
\\&-\xi\T \left(I_n\otimes(\mathcal{C}\T \left[
\begin{array}{cc} 0&0\\
0& I_{p_2}\end{array} \right]\mathcal{C})\right)\xi
\\&~+\frac{ 2\lambda_{\max}(P_k) \lambda_{\max}(\mathcal{P}^{-1})
\|\mathcal{L}_{\varepsilon}\|}{\varepsilon\lambda_{\min}(P_k) \lambda_{\min}(\mathcal{P}^{-1}) }V_k
\\ \leq &-\xi\T\left(P_k\otimes (\mathcal{P}^{-1}\mathcal{P}^{-1})\right)\xi,
\\&~+\frac{ 2\lambda_{\max}(P_k) \lambda_{\max}(\mathcal{P}^{-1}) \|\mathcal{L}_{\varepsilon}\|}{\varepsilon\lambda_{\min}(P_k) \lambda_{\min}(\mathcal{P}^{-1}) }V_k
\\ \leq &-\left(\frac{\lambda_{\min}(\mathcal{P}^{-1})}{\varepsilon}-\frac{ 2\lambda_{\max}(P_k) \lambda_{\max}(\mathcal{P}^{-1}) \|\mathcal{L}_{\varepsilon}\|}{\varepsilon\lambda_{\min}(P_k) \lambda_{\min}(\mathcal{P}^{-1}) }\right)V_k,
\end{align*}
where we have used \eqref{eq:P-full} and the fact that $\theta\leq \beta_k$, $ k\in \Gamma_c$.
It then follows that $\dot V_k \leq -\frac{1}{\varepsilon}\lambda_k V_k$, $\forall k\in \Gamma_c$,
if $\|\mathcal{L}_{\varepsilon}\|<\frac{\lambda_{\min}(P_k) \lambda_{\min}(\mathcal{P})}{4\lambda_{\max}(P_k) \lambda_{\max}^2(\mathcal{P}) }$, where $\lambda_k=\frac{1}{2\lambda_{\max}(\mathcal{P})}$, $\forall k\in \Gamma_c$.

On the other hand, for all $k\in \Gamma_d$, we have that
\begin{align*}
\dot V_k\leq & ~2\xi\T\left(P_k\otimes(\mathcal{P}^{-1}\mathcal{A})\right)\xi
+ 2\xi\T\left(P_k\otimes \mathcal{P}^{-1}\right)\mathcal{L}_{\varepsilon}\xi
\\& -2\xi\T\left(P_k\otimes\left(\mathcal{C}\T \left[
\begin{array}{cc} I_{p_1}&0\\
0& 0\end{array} \right] \mathcal{C}\right)\right)\xi
\\&-2\xi\T\left(P_kL_k\otimes\left(\mathcal{C}\T \left[
\begin{array}{cc} 0&0\\
0& I_{p_2}\end{array} \right] \mathcal{C}\right)\right)\xi
\\ \leq &~ \xi\T\left(P_k\otimes(\mathcal{P}^{-1}(\mathcal{A}\mathcal{P}+\mathcal{P}\mathcal{A}\T)\mathcal{P}^{-1})
\right)\xi
\\&~+2\lambda_{\max}(P_k) \lambda_{\max}(\mathcal{P}^{-1}) \|\mathcal{L}_{\varepsilon}\| \|\xi\|^2
\\ \leq &~2 \xi\T\left(P_k\otimes \left(\mathcal{C}\T\left[
\begin{array}{cc} I_{p_1}&0\\
0& \theta I_{p_2}\end{array} \right] \mathcal{C}\right)\right)\xi
-\frac{ \lambda_{\min}(\mathcal{P}^{-1})}{\varepsilon}V_k
\\&~+\frac{ 2\lambda_{\max}(P_k) \lambda_{\max}(\mathcal{P}^{-1}) \|\mathcal{L}_{\varepsilon}\|}{\varepsilon\lambda_{\min}(P_k) \lambda_{\min}(\mathcal{P}^{-1}) }V_k,
  \end{align*}
where we have used \eqref{eq:P-full}. Note that
$\lambda_{\max}\left(\mathcal{C}\T \left[
\begin{array}{cc} I_{p_1}&0\\
0& \theta I_{p_2}\end{array} \right]\mathcal{C}\right)\\=\max\{\theta,1\}$.
It follows that $\dot V_k \leq \frac{1}{\varepsilon}\lambda_k V_k$, $\forall k\in \Gamma_d$, if $ \|\mathcal{L}_{\varepsilon}\|<\frac{\lambda_{\min}(P_k) \lambda_{\min}(\mathcal{P})}{2\lambda_{\max}(P_k) \lambda_{\max}^2(\mathcal{P}) }$,
where $\lambda_k=2\max\{\theta,1\}\lambda_{\max}(\mathcal{P})$, $\forall k\in \Gamma_d$.

Following the similar analysis of \cite{Liberzon_ICSM1999,ZhaiGuisheng_AJC00},
we let
$\sigma=p_{j}$ on $[t_{j-1},t_j)$ for $p_{j}\in \Gamma$. Then, for any $t$ satisfying
$t_0<t_1<\dots<t_\ell<t<t_{\ell+1}$, define $V=\varepsilon\xi\T (P_{\sigma(t)}
\otimes \mathcal{P}^{-1})\xi$ for \eqref{eq:switch-full}.
We have that, $\forall\zeta\in [t_{j-1},t_j)$,
\begin{subequations}
  \begin{align*}
V(\zeta)\leq & ~e^{-\frac{1}{\varepsilon}\lambda_{p_j}(\zeta-t_{j-1})} V(t_{j-1})
\\ \leq &  ~e^{-\frac{1}{\varepsilon}\lambda^c(\zeta-t_{j-1})} V(t_{j-1}), \qquad p_{j}\in \Gamma_c,
  \end{align*}
    \begin{align*}
V(\zeta) \leq & ~ e^{\frac{1}{\varepsilon}\lambda_{p_j}(\zeta-t_{j-1})} V(t_{j-1})
\\ \leq & ~e^{\frac{1}{\varepsilon}\lambda^d(\zeta-t_{j-1})} V(t_{j-1}), \qquad p_{j}\in \Gamma_d,
  \end{align*}
\end{subequations}
where $\lambda^c=\min_{k\in \Gamma_c}\lambda_k=
\frac{1}{2\lambda_{\max}(\mathcal{P})}$,
$\lambda^d=\max_{k\in \Gamma_d}\lambda_k=
2\max\{\theta,1\}\lambda_{\max}(\mathcal{P})$.
Define $a=\frac{\lambda_{\max}(\mathcal{P})}{\lambda_{\min}(\mathcal{P})}
\max_{k,j\in \Gamma}\frac{\lambda_{\max}(P_k)}{\lambda_{\min}(P_j)}$.
We then know that $V(t_j)\leq a \lim_{t\uparrow t_j}V(t)$. Thus, it follows that
\begin{equation*}
V(t)\leq a^{\rho}e^{\frac{1}{\varepsilon}\lambda^dT^d_{\overline{t}_0}(t)-\frac{1}{\varepsilon}\lambda^cT^c_{\overline{t}_0}(t)} V(\overline{t}_0),
\end{equation*}
where $\rho$ denotes times of switching during $[\overline{t}_0,t)$. Note that $\rho\leq \frac{t-\overline{t}_0}{\tau_d}$.
Given that $\kappa\geq\kappa^*=\frac{\lambda^d+\lambda}{\lambda^c-\lambda}$, for some $\lambda\in(0,\lambda^c)$, it follows from Assumption \ref{assm:output-time} that
$T^c_{\overline{t}_0}(t)\geq \kappa^* T^d_{\overline{t}_0}(t)$ for all $t\geq \overline{t}_0$. This implies that $\lambda^dT^d_{\overline{t}_0}(t)-\lambda^cT^c_{\overline{t}_0}(t)\leq -\lambda(T^d_{\overline{t}_0}(t)+T^c_{\overline{t}_0}(t))$, for all $t\geq \overline{t}_0$ and we therefore know that
 \begin{align*}
V(t)\leq &~ a^{\rho}e^{-\frac{1}{\varepsilon}\lambda(t-\overline{t}_0)} V(\overline{t}_0)
\\ \leq& ~e^{\frac{t-\overline{t}_0}{\tau_d}\ln a-\frac{1}{\varepsilon}\lambda(t-\overline{t}_0)} V(\overline{t}_0)
\\ =&~e^{-\left(\frac{1}{\varepsilon}\lambda-\frac{\ln a}{\tau_d}\right)(t-\overline{t}_0)} V(\overline{t}_0).
\end{align*}

Furthermore, set $\lambda=\alpha\lambda^c$, where some $\alpha\in (0,1)$.
We then have that $\kappa^*=\frac{\alpha+4\max\{\theta,1\}\lambda^2_{\max}(\mathcal{P})}{1-\alpha}$, and
\begin{align*}
V(t)\leq  ~e^{-\left(\frac{\alpha}{2\varepsilon\lambda_{\max}(\mathcal{P})}-\frac{\ln a}{\tau_d}\right)(t-\overline{t}_0)} V(\overline{t}_0).
\end{align*}

It follows that if $\varepsilon<\frac{\alpha\tau_d}{2\lambda_{\max}(\mathcal{P})\ln a}$,
we have for \eqref{eq:switch-full} that
\begin{align*}
\|\xi(t)\|\leq c^*e^{-\frac{1}{2}\left(\frac{\alpha}{2\varepsilon\lambda_{\max}(\mathcal{P})}-\frac{\ln a}{\tau_d}\right)(t-\overline{t}_0)}\|\xi(\overline{t}_0)\|,
\end{align*}
where $c^*=\sqrt{\frac{\lambda_{\max}(\mathcal{P})\max_{k\in\Gamma}\lambda_{\max}(P_k)}
{\lambda_{\min}(\mathcal{P})\min_{k\in\Gamma}\lambda_{\min}(P_k)}}$.

Therefore, we choose $\varepsilon^*$ satisfying $\varepsilon^*<\frac{\alpha\tau_d}{2\lambda_{\max}(\mathcal{P})\ln a}$ and
$ \|\mathcal{L}_{\varepsilon^*}\|<\min_{k\in\Gamma}\frac{\lambda_{\min}(P_k) \lambda_{\min}(\mathcal{P}) }{4\lambda_{\max}(P_k) \lambda_{\max}^2(\mathcal{P}) }$.
It then follows
that $\lim_{t\rightarrow\infty}(\chi_{i}(t)-\widehat{\chi}_{i}(t))=0$, $i=1,2\ldots,n$.
\endproof

\begin{remark}
Note that the condition of $\kappa$ is necessary when the communication topology is switching.
Roughly speaking, we need to guarantee that the influence of ``the good topology''
beats that of ``the bad topology'' since the states of open-loop systems might diverge very fast
due to the existence of unstable modes.
The parameter $\kappa$ is used to describe the relationship between $T^c_{\overline{t}_0}(t)$ and $T^d_{\overline{t}_0}(t)$,
{\em i.e.}, the remaining times of ``good topology'' and ``bad topology'', respectively.
The derived upper bound on $\kappa$ might not be tight.
However, we would like to emphasize that the
significance is on the qualitative effects instead of quantitative effects.
In practical applications, we can use an empirical approach to derive a feasible $\kappa$,
as illustrated in Section \ref{sec:simulation}.
\end{remark}

From the unified observer design, we then have that
\begin{equation}
\widehat{\overline{x}}_i=(T_i\T T_i)^{-1}T_i\T \widehat{\chi}_i
=[\widehat{\overline{x}}_{i1}\T,\widehat{\overline{x}}_{i2}\T]\T, ~i=1,2,\dots,n,\label{eq:output-x-hat}
\end{equation}
which will be used in the control input design.

\subsection{Main Results}

In this section, we show that the observer architecture introduced in the previous sections
provide an asymptotically stable closed-loop system, as presented in Theorems
\ref{thm:output} below.
The observer-based controller is proposed as
\begin{equation}
u_i=F_i\widehat{\overline{x}}_{i1}+(\Gamma_{i}-F_i\Pi_{i})
\widehat{\overline{x}}_{i2},\label{eq:control-full}
\end{equation}
where $\Pi_{i}$ and $\Gamma_{i}$ are the solutions of
the regulator equation \eqref{eq:output-regulator}, and $\widehat{\overline{x}}_{i1}$ and $\widehat{\overline{x}}_{i2}$ can be obtained from \eqref{eq:observer-full}.

\begin{thm}\label{thm:output}
Let Assumptions \ref{assm:output-dwell}, \ref{assm:output-time},
\ref{assm:A-C} and \ref{assm:A-B} hold and
assume that $\kappa\geq\frac{\alpha+4\max\{1,\theta\}\lambda^2_{\max}(\mathcal{P})}{1-\alpha}$,
where $\alpha\in (0,1)$, $\theta$ and $\mathcal{P}$ are
given by \eqref{eq:P-full}.
Then, there exists $\varepsilon^*\in (0,1]$ such that,
if $\varepsilon\in (0,\varepsilon^*]$, \eqref{eq:control-full} ensures
that $\lim_{t\rightarrow\infty}e_i(t)=0$, $i=1,2\ldots,n$,
for the multi-agent system \eqref{eq:state}-\eqref{eq:output-error}.
\end{thm}

\proof
Follows from Lemmas \ref{lem:control-full} and \ref{lem:observer-full}, and the separation principle.
\endproof

\begin{remark}
If the leader-follower communication topology $\overline{G}$ is time-invariant, Assumptions \ref{assm:output-dwell} and \ref{assm:output-time} are not needed, and therefore the high-gain parameter only depends on the non-identical dynamics of the agents.
\end{remark}



\section{Case Studies}\label{sec:case-study}

We notice that \eqref{eq:observer-full} give a unified way using
$y_{si}$ and $\zeta_i$ to
estimate $x_i$, $\omega_i$, and $x_0$. One drawback of such a general approach
is that the dimension of the observer $\widehat{\chi}_i$ may be unnecessarily large
for some cases with special structures. We next give particular structural designs on
two special cases, {\em i.e.}, the case
when $(A_i,C_{si})$ is observable and the case when $(A_i,C_{di})$ is observable\footnote{These two cases are
special cases of the first item of Assumption \ref{assm:A-C}.}.

\subsection{Case I: $(A_i,C_{si})$ is observable}\label{sec:cor1}

\begin{figure}
\begin{center}
\includegraphics[scale=0.45]{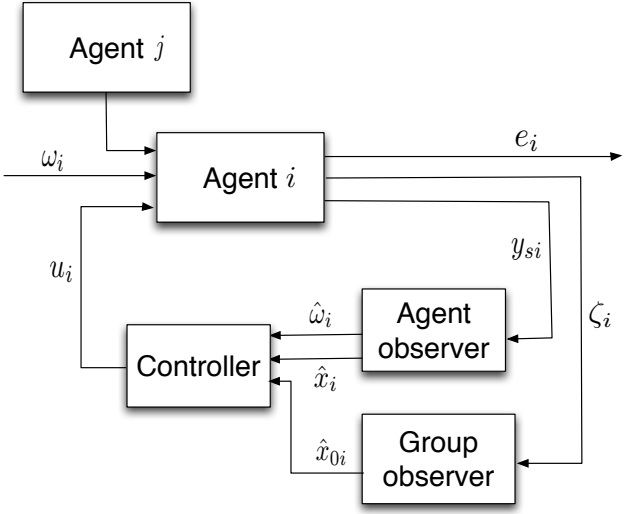}
  \caption{Control architecture for agent $\nu_i$}
\label{fig:architecture1}
 \end{center}
 \end{figure}

In this section, we use $y_{si}$ to estimate both
$x_i$ and $\omega_i$ and use $\zeta_i$ to estimate $x_0$.
The control of each agent has the structure shown in
Fig. \ref{fig:architecture1}.

We replace the first item of Assumption \ref{assm:A-C} with that
$(A_i,C_{si})$, for all $i=1,2,\dots,n$ is observable.

\textbf{Step I: redundant mode remove}

We first write the state and output of $x_i$ and $\omega_i$ for each agent
in the compact form
\begin{subequations}
\begin{align*}
\left[\begin{array}{c} \dot x_i\\ \dot \omega_i \end{array} \right]
&=\left[\begin{array}{cc} A_i & 0\\ 0& S_i \end{array} \right]
\left[\begin{array}{c} x_i\\ \omega_i \end{array} \right]
+\left[\begin{array}{c} B_i\\ 0 \end{array} \right]u_i,
\\
 y_{si}&=\left[\begin{array}{cc} C_{si} & C_{wi} \end{array} \right]
\left[\begin{array}{c} x_i\\ \omega_i \end{array} \right].
\end{align*}
\end{subequations}
We can then construct a new state
$\overline{x}_i=W_i\left[\begin{array}{c}
x_i\\ \omega_i\end{array} \right]$ and perform
the state transformation
such that
\begin{subequations}
\begin{align*}
\dot {\overline{x}}_i&=\overline{A}_i\overline{x}_i+\overline{B}_iu_i=
\left[\begin{array}{cc} A_i & \overline{A}_{i12}\\ 0& \overline{A}_{i22} \end{array} \right]\overline{x}_i
+\left[\begin{array}{c} B_i\\ 0\end{array} \right]u_i,
\\
 y_{si}&=\overline{C}_{i}\overline{x}_i=\left[\begin{array}{cc} C_{si} & \overline{C}_{i12}
 \end{array} \right]\overline{x}_i.
\end{align*}
\end{subequations}
Similar to Section \ref{sec:remove1}, we can show that the pair
$(\overline{A}_i,\overline{C}_i)$ is observable
and the eigenvalues of $\overline{A}_{i22}$ are a subset of the eigenvalues
of $S_i$, $i=1,2,\dots,n$.

\textbf{Step II: agent observer}

Based on the information of the individual output information $y_{si}$,
the following individual observer for each agent $\nu_i$ is proposed
\begin{subequations}\label{eq:x-w-hat-special1}
\begin{equation}
\dot{\widehat {\overline{x}}}_i=\overline{A}_i\widehat{\overline{x}}_i
+\overline{B}_iu_i+K_{ai}\left(\overline{C}_{i}\widehat{\overline{x}}_i-y_{si}\right),
\end{equation}
\begin{equation}
[\widehat{x}_i\T,\widehat{\omega}_{i}\T]\T=W_i^{-1}\widehat{\overline{x}}_i, ~i=1,2,\dots,n,
\end{equation}
\end{subequations}
where $K_{ai}$ is chosen such that $\overline{A}_i+K_{ai}\overline{C}_{i}$
is Hurwitz stable, $i=1,2\ldots,n$.

\textbf{Step III: group observer}

We transform \eqref{eq:leader} into its canonical form.
Define $\chi_0=T_0x_0\in\mathbb{R}^{pn_0}$, where
\begin{equation*}
T_0=\left[
\begin{array}{c} C_0 \\ \vdots \\ C_0A_0^{n_0-1} \end{array} \right].
\end{equation*}
Therefore, it follows that
\begin{subequations}
  \begin{align*}
\dot\chi_0&=(\mathcal{A}_0+\mathcal{L}_0)\chi_0,
\\
y_0&=\mathcal{C}_0\chi_0,
  \end{align*}
\end{subequations}
where $\mathcal{A}_0=\begin{bmatrix} 0 & I_{p(n_0-1)}\\ 0 & 0 \end{bmatrix}\in \mathbb{R}^{pn_0\times pn_0}$,
$\mathcal{L}_0=\\ \left[
\begin{array}{c} 0 \\   C_0A_0^{n_0}(T_0\T T_0)^{-1}T_0\T \end{array} \right]$,
$\mathcal{C}_0=\left[
\begin{array}{cc} I_p & 0 \end{array} \right]\in \mathbb{R}^{p\times pn_0}$.

Then, based on the neighbor-based group output information $\zeta_i$, the distributed observer is proposed
\begin{subequations}\label{eq:observer-x0-special1}
\begin{align}
\dot{\widehat \chi}_{0i}=~(\mathcal{A}_0+\mathcal{L}_0)\widehat \chi_{0i}&-S(\varepsilon)\mathcal{P}\mathcal{C}_0\T
\left(\sum_{j=0}^na_{ij}(t)(y_{di}-y_{dj})\right.\notag\\& \left.-\sum_{j=0}^na_{ij}(t)(\widehat{y}_{i}-\widehat{y}_{j})\right),
\end{align}
  \begin{align}
\widehat x_{0i}=(T_0\T T_0)^{-1}T_0\T\widehat \chi_{0i},~~~
 i=1,2\ldots,n,
  \end{align}
\end{subequations}
where $a_{ij}(t)$, $i=0,1,\ldots,n$, $j=0,1,\ldots,n$, is
entry $(i,j)$ of the adjacency matrix $\overline{{A}}_{\sigma(t)}$ associated with $\overline{{G}}_{\sigma(t)}$ defined
in Section \ref{sec:graph} at time $t$, the relative estimation information
$\sum_{j=0}^na_{ij}(t)(\widehat{y}_{i}-\widehat{y}_{j})$ is obtained using the communication infrastructure with
$\widehat{y}_i=C_{di}\widehat x_i-\mathcal{C}_0\widehat \chi_{0i}$, $i=1,2,\dots,n$ and $\widehat{y}_0=0$. In addition, $S(\varepsilon)=\diag(I_p\varepsilon^{-1},
I_p\varepsilon^{-2},\dots,I_p\varepsilon^{-n_0} )$, where $\varepsilon\in (0,1]$
is a positive constant,
and $\mathcal{P}=\mathcal{P}\T$ is a positive definite matrix satisfying
\begin{align}
\mathcal{A}_0\mathcal{P}+\mathcal{P}\mathcal{A}_0\T-2\theta\mathcal{P}\mathcal{C}_0\T \mathcal{C}_0 \mathcal{P}+ I_{pn_0}=0,\label{eq:P-special1}
\end{align}
and $\theta$ is a positive constant satisfying $\theta<\frac{1}{2}\min_{\overline{{G}}_{k}\in \overline{\mathbb{G}}_c}\\ \min\Re\{\lambda(L_k)\}$.

\textbf{Step IV: controller design}

The observer-based controller is proposed as
\begin{equation}
u_i=F_i\widehat{x}_i+(\Gamma_{1i}-F_i\Pi_{1i})\widehat{\omega}_i + (\Gamma_{2i}-F_i\Pi_{2i})\widehat{x}_{0i},\label{eq:control-special1}
\end{equation}
where $\Pi_{1i}$, $\Gamma_{1i}$, $\Pi_{2i}$, and $\Gamma_{2i}$ are the solutions of the following regulator equations
\begin{subequations}\label{eq:regulator-cor1}
  \begin{align}
\Pi_{1i}S_i&=A_i\Pi_{1i}+B_i\Gamma_{1i}, \\
0&=D_{si}\Pi_{1i}+D_{wi},  \\
\Pi_{2i}A_0&=A_i\Pi_{2i}+B_i\Gamma_{2i}, \\
0&=D_{si}\Pi_{2i}+D_{0},\quad i=1,2\dots,n,
  \end{align}
\end{subequations}
and $F_i$ is chosen such that $A_i+B_iF_i$ is Hurwitz.

\begin{cor}\label{thm:special1}
Let Assumptions \ref{assm:output-dwell}, \ref{assm:output-time},
\ref{assm:A-C} (the first item is replaced by that
$(A_i,C_{si})$ is observable), and \ref{assm:A-B} hold and
assume that $\kappa\geq\frac{\alpha+4\theta\lambda^2_{\max}(\mathcal{P})}{1-\alpha}$,
where $\alpha\in (0,1)$, $\theta$ and $\mathcal{P}$ are given by \eqref{eq:P-special1}.
Also, let $\widehat{x}_i$ and $\widehat{\omega}_i$ be obtained in
\eqref{eq:x-w-hat-special1}, and $\widehat{x}_{0i}$ be obtained in
\eqref{eq:observer-x0-special1}. Then, there exists $\varepsilon_1^*\in (0,1]$ such that,
if $\varepsilon\in (0,\varepsilon_1^*]$, \eqref{eq:control-special1} ensures that
$\lim_{t\rightarrow\infty}e_i(t)=0$, $i=1,2\ldots,n$, for the multi-agent system \eqref{eq:state}-\eqref{eq:output-error}.
\end{cor}
\proof
The proof is straightforward following the similar analysis given in Lemmas \ref{lem:control-full} and \ref{lem:observer-full}.
\endproof

\subsection{Case II: $(A_i,C_{di})$ is observable}\label{sec:cor2}

\begin{figure}
\begin{center}
\includegraphics[scale=0.45]{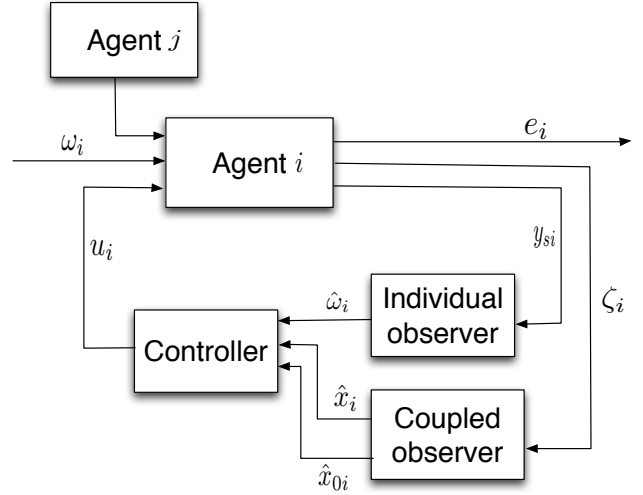}
  \caption{Control architecture for agent $\nu_i$}
\label{fig:architecture2}
 \end{center}
 \end{figure}

In this section, we use $y_{si}$ to estimate $\omega_i$ and use $\zeta_i$ to
estimate both $x_i$ and $x_0$.
The control of each agent is supposed to have the structure shown in Fig.
\ref{fig:architecture2}.

We also replace the first item of Assumption \ref{assm:A-C} with that
$(A_i,C_{di})$, for all $i=1,2,\dots,n$ is observable.

\textbf{Step I: redundant mode remove}

We first write the state and output of $x_i$ and $x_0$ for each agent
in the compact form
\begin{subequations}
\begin{align*}
\left[\begin{array}{c} \dot x_i\\ \dot x_0 \end{array} \right]
=\left[\begin{array}{cc} A_i & 0\\ 0& A_0 \end{array} \right]
\left[\begin{array}{c} x_i\\ x_0 \end{array} \right]
+\left[\begin{array}{c} B_i\\ 0 \end{array} \right]u_i,
\\
 e_{di}=y_{di}-y_{d0}=\left[\begin{array}{cc} C_{di} & -C_{0} \end{array} \right]
\left[\begin{array}{c} x_i\\ x_0  \end{array} \right].
\end{align*}
\end{subequations}
We can then construct a new state
$\overline{x}_i=W_i\left[\begin{array}{c}
x_i\\ x_0\end{array} \right]$ and perform
the state transformation
such that
\begin{subequations}
\begin{align*}
\dot {\overline{x}}_i=\overline{A}_i\overline{x}_i+\overline{B}_iu_i=
\left[\begin{array}{cc} A_i & \overline{A}_{i12}\\ 0& \overline{A}_{i22} \end{array} \right]\overline{x}_i
+\left[\begin{array}{c} B_i\\ 0\end{array} \right]u_i,
\\
 e_{di}=\overline{C}_{i}\overline{x}_i=\left[\begin{array}{cc} C_{di} & \overline{C}_{i12}
 \end{array} \right]\overline{x}_i.
\end{align*}
\end{subequations}
Similarly, we can show that pair $(\overline{A}_i,\overline{C}_i)$ is observable
and the eigenvalues of $\overline{A}_{i22}$ are a subset of the eigenvalues
of $A_0$, $i=1,2,\dots,n$.

\textbf{Step II: coupled observer}

We next define $\chi_i=T_i\overline{x}_i\in\mathbb{R}^{p\overline{n}}$, $i=1,2,\dots,n$, where $\overline{n}=n_0+\max_{i=1,2,\dots,n}n_i$, and
\begin{equation*}
T_i=\left[
\begin{array}{c} \overline{C}_i \\ \vdots \\ \overline{C}_i\overline{A}_i^{\overline{n}-1} \end{array} \right].
\end{equation*}
Therefore, it follows that
\begin{subequations}\label{eq:close-special2}
  \begin{equation}
\dot\chi_i=(\mathcal{A}+\mathcal{L}_i)\chi_i+\mathcal{B}_iu_i,
  \end{equation}
  \begin{equation}
e_{di}=\mathcal{C}\chi_i,\quad i=1,2,\dots,n,
  \end{equation}
\end{subequations}
where $\mathcal{A}=\left[
\begin{array}{cc} 0 & I_{p(\overline{n}-1)}\\ 0 & 0 \end{array} \right]\in \mathbb{R}^{p\overline{n}\times p\overline{n}}$,
$\mathcal{L}_i=\left[
\begin{array}{c} 0 \\  L_i \end{array} \right]$, 
$\mathcal{B}_i=T_i\overline{B}_i$, $\mathcal{C}=\left[
\begin{array}{cc} I_p & 0 \end{array} \right]\in \mathbb{R}^{p\times p\overline{n}}$ for some matrix $L_i\in\mathbb{R}^{p\times p\overline{n}}$.

Based on the neighbor-based group output information $\zeta_i$, the distributed observer is proposed for \eqref{eq:close-special2} as
\begin{subequations}\label{eq:observer-special2}
\begin{align}
\dot{\widehat{\chi}}_i=&(\mathcal{A}+\mathcal{L}_i)\widehat{\chi}_i+\mathcal{B}_iu_i+S(\varepsilon)
\mathcal{P}\mathcal{C}\T
\notag\\ &\times\left(\!\sum_{j=0}^na_{ij}(t)(y_{di}-y_{dj})-\sum_{j=0}^na_{ij}(t)(\widehat{y}_{i}-\widehat{y}_{j})\!\!\right),
\end{align}
\begin{align}
[\widehat{x}_i\T,\widehat{x}_{0i}\T]\T=W_i^{-1}(T_i\T T_i)^{-1}T_i\T
\widehat{\chi}_i, ~i=1,2,\dots,n,
\end{align}
\end{subequations}
where $a_{ij}(t)$, $i=0,1,\ldots,n$, $j=0,1,\ldots,n$, is
entry $(i,j)$ of the adjacency matrix $\overline{{A}}_{\sigma(t)}$ associated with $\overline{{G}}_{\sigma(t)}$ defined
in Section \ref{sec:graph} at time $t$, $\widehat{y}_i=\mathcal{C}\widehat{\chi}_i$,
$i=1,2,\dots,n$, $\widehat{y}_0=0$.
In addition, $S(\varepsilon)=\diag(I_p\varepsilon^{-1},I_p\varepsilon^{-2},
\dots,I_p\varepsilon^{-\overline{n}} )$, where $\varepsilon\in (0,1]$
is a positive constant, and $\mathcal{P}=\mathcal{P}\T$ is a positive definite matrix satisfying
\begin{align}
\mathcal{A}\mathcal{P}+\mathcal{P}\mathcal{A}\T-2\theta \mathcal{P}\mathcal{C}\T\mathcal{C} \mathcal{P}+I_{p\overline{n}}=0,\label{eq:P-special2}
\end{align}
where $\theta$ is a positive constant satisfying $\theta<\frac{1}{2}\min_{\overline{{G}}_{k}\in \overline{\mathbb{G}}_c}\\ \min\Re\{\lambda(L_k)\}$.

\textbf{Step III: individual observer}

Based on the information of $\widehat{x}_i$ and the individual output information $y_{si}$,
the following individual observer for each agent is proposed
\begin{equation}
\dot{\widehat{\omega}}_i=S_i\widehat{\omega}_i +
K_{si}\left(C_{si}\widehat{x}_i+C_{wi}\widehat{\omega}_i-y_{si}\right),~
i=1,2\ldots,n,\label{eq:w-hat-special2}
\end{equation}
where $K_{si}$ is chosen such that $S_i+K_{si}C_{wi}$ is Hurwitz stable.

\textbf{Step IV: controller design}

The observer-based controller is proposed as
\begin{equation}
u_i=F_i\widehat{x}_i+(\Gamma_{1i}-F_i\Pi_{1i})\widehat{\omega}_i + (\Gamma_{2i}-F_i\Pi_{2i})\widehat{x}_{0i},\label{eq:control-special2}
\end{equation}
where $\Pi_{1i}$, $\Gamma_{1i}$, $\Pi_{2i}$, and $\Gamma_{2i}$ are the solutions of the following regulator equations
\begin{subequations}\label{eq:regulator-cor2}
  \begin{align}
\Pi_{1i}S_i&=A_i\Pi_{1i}+B_i\Gamma_{1i}, \\
0&=D_{si}\Pi_{1i}+D_{wi},  \\
\Pi_{2i}A_0&=A_i\Pi_{2i}+B_i\Gamma_{2i}, \\
0&=D_{si}\Pi_{2i}+D_{0},\quad i=1,2\dots,n.
  \end{align}
\end{subequations}
and $F_i$ is chosen such that $A_i+B_iF_i$ is Hurwitz.

\begin{cor}\label{thm:special2}
Let Assumptions \ref{assm:output-dwell}, \ref{assm:output-time}, \ref{assm:A-C} (the first item is replaced by that
$(A_i,C_{di})$ is observable), and
\ref{assm:A-B} hold and
assume that $\kappa\geq\frac{\alpha+4\theta\lambda^2_{\max}(\mathcal{P})}{1-\alpha}$, where $\alpha\in (0,1)$, $\theta$ and $\mathcal{P}$ are given by \eqref{eq:P-special2}. Also, let $\widehat{x}_i$ and $\widehat{x}_{0i}$ be obtained in \eqref{eq:observer-special2}, and $\widehat{\omega}_i$ be obtained in \eqref{eq:w-hat-special2}. Then, there exists $\varepsilon_2^*\in (0,1]$ such that, if $\varepsilon\in (0,\varepsilon_2^*]$, \eqref{eq:control-special2} ensures that $\lim_{t\rightarrow\infty}e_i(t)=0$, $i=1,2\ldots,n$, for the multi-agent system \eqref{eq:state}-\eqref{eq:output-error}.
\end{cor}
\proof
See \cite{MengZiyang_CDC13}.
\endproof
\section{Simulation Results}
\label{sec:simulation}
In this section, we illustrate the theoretical results.
Consider a network of three agents as shown in Fig. \ref{fig:COMMTOP1}. We assume that
the adjacency matrix $\overline{{A}}_{\sigma(t)}$ associated with
$\overline{{G}}_{\sigma(t)}$ is switching periodically. Denote $\ell=0,20,40\dots$.

$ \overline{{A}}=\begin{cases}
\left[\begin{matrix}0& 0 & 0& 0 \\ 1& 0 & 1 &0
\\ 0& 1 &0 &0\\ 0& 0&1&0 \end{matrix}\right], & {\rm when}~~ t\in[\ell,\ell+6), \\
\left[\begin{matrix}0& 0 & 0& 0 \\ 1& 0 & 0 &0
\\ 0& 1 &0 &1\\ 0& 0&1&0 \end{matrix}\right], & {\rm when}~~ t\in[\ell+6,\ell+12), \\
\left[\begin{matrix}0& 0 & 0& 0 \\ 1& 0 & 0 &0
\\ 0& 1 &0 &0\\ 0& 0&1&0 \end{matrix}\right], & {\rm when}~~ t\in[\ell+12,\ell+18), \\
\left[\begin{matrix}0& 0 & 0& 0 \\0& 0 & 0& 0
\\ 0& 0 &0& 0 \\ 0& 0&0& 0 \end{matrix}\right], & {\rm when}~~ t\in[\ell+18,\ell+20).
\end{cases}\notag$

\textbf{Example 1}

We give an example to validate Theorem \ref{thm:output},
the dynamics of the agents are described as
$A_1=\left[\begin{array}{ccc}0 & 3 & 0\\
0 & 0 & 2\\ 0 & -1 & 0 \end{array}\right]$, $B_1=\left[\begin{array}{c}0 \\ 0 \\ 1\end{array}\right]$, $C_{s1}=C_{d1}=D_{s1}=\left[\begin{array}{ccc}1 & 1 & 1\end{array}\right]$,
$A_2=\left[\begin{array}{cc}1 & 0 \\
0 & 0 \end{array}\right]$, $B_2=\left[\begin{array}{cc}1 \\ 1\end{array}\right]$,
$C_{s2}=\left[\begin{array}{cc}1 & 0\end{array}\right]$,
$C_{d2}=\left[\begin{array}{cc}0 & 1\end{array}\right]$,
$D_{s2}=\left[\begin{array}{cc}1 & 1\end{array}\right]$,
$A_3=\left[\begin{array}{cc}0 & 1 \\
 -2& -2 \end{array}\right]$, $B_3=\left[\begin{array}{c}0 \\ 1\end{array}\right]$,
 $C_{s3}=C_{d3}=D_{s3}=\left[\begin{array}{ccc}1 & 0\end{array}\right]$.
The dynamics of the individual autonomous exosystems are described as $S_i=0$,
$C_{wi}=D_{wi}=-1$, $i=1,2,3$, and $\omega_1(0)=-2$, $\omega_2(0)=-4$, and $\omega_3(0)=-6$.
The dynamics of the group autonomous exosystem are described as $A_0=\left[\begin{array}{cc}0 & 1\\
-1 & 0  \end{array}\right]$,
 $C_0=\left[\begin{array}{cc}1 & 0 \end{array}\right]$, $D_{0}=-C_0$.

Following the design scheme proposed in Section \ref{sec:unified},
for the solutions of regulator
equations \eqref{eq:output-regulator}, we have that $F_1=\left[\begin{array}{ccc} -1 & -4.5 & -6 \end{array}\right]$,
 $\Pi_{1}=\left[\begin{array}{ccc}1& 1.0345 & -0.4138\\
0& 0.1379 & 0.3448\\0& -0.1724 & 0.0690  \end{array}\right]$,
 $\Gamma_{1}=\left[\begin{array}{ccc}0 & 0.0690 & 0.1724 \end{array}\right]$ for agent $\nu_1$,
$F_2=\left[\begin{array}{cc} -2 & -6 \end{array}\right]$,
$\Pi_{2}=\left[\begin{array}{ccc}0& 0.4&-0.2 \\ 1&0.6&0.2 \end{array}\right]$,
$\Gamma_{2}=\left[\begin{array}{ccc}0& -0.2 & 0.6 \end{array}\right]$ for agent $\nu_2$,
  $F_3=\left[\begin{array}{cc} 0 & -1 \end{array}\right]$,
$\Pi_{3}=\left[\begin{array}{ccc}1& 1 & 0\\
0& 0 & 1  \end{array}\right]$,
 $\Gamma_{3}=\left[\begin{array}{ccc}2& 1 & 2 \end{array}\right]$ for agent $\nu_3$.
We also have $\varepsilon=0.2$ for \eqref{eq:observer-full} and
$\theta=0.1$ for \eqref{eq:P-full}.
\begin{figure}
\begin{center}
\includegraphics[width=9cm,height=7cm]{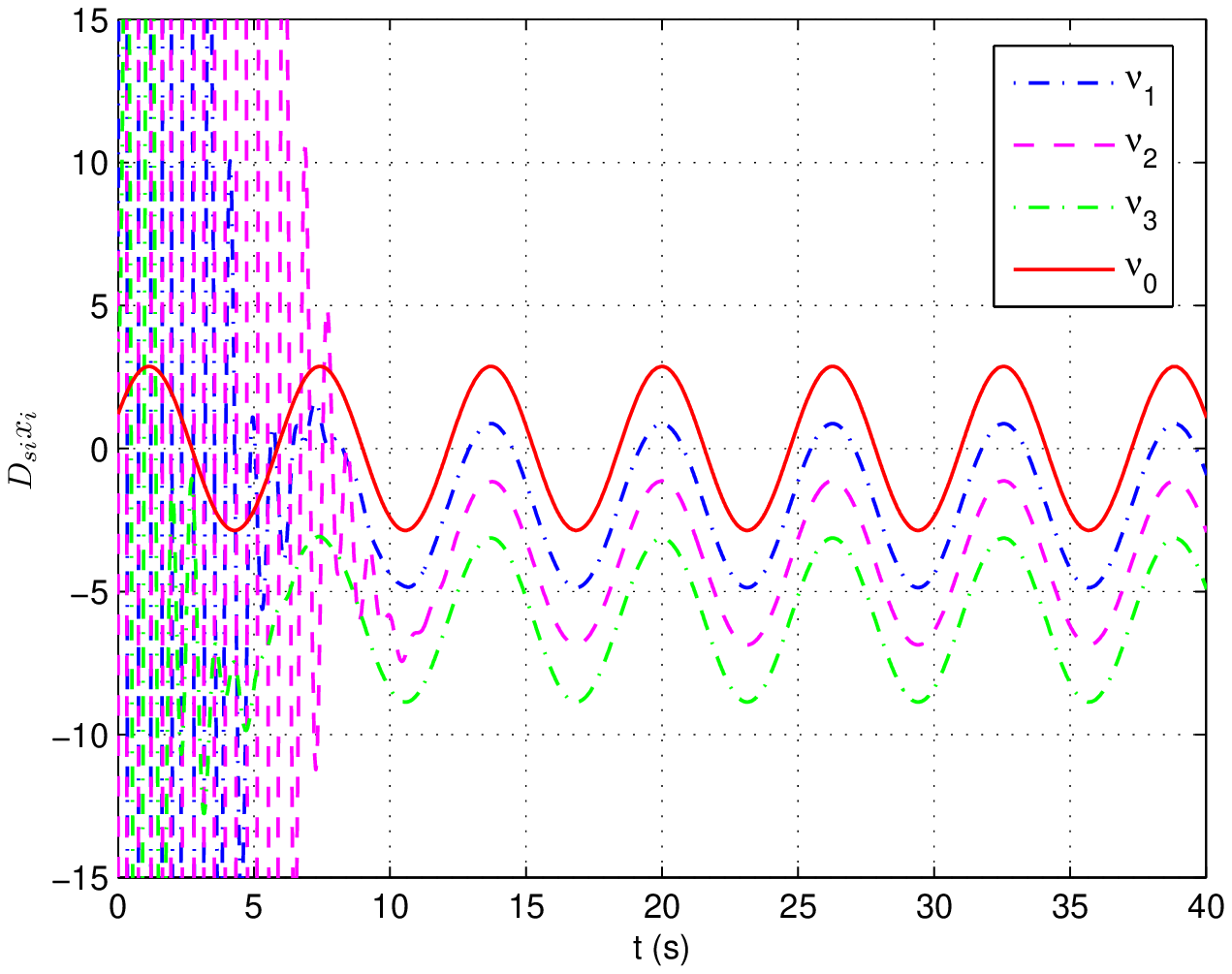}
  \caption{Output convergence of system \eqref{eq:state}, \eqref{eq:disturbance},
and \eqref{eq:leader}
  under the observer-based controller \eqref{eq:control-full}
  for Theorem \ref{thm:output}}
\label{fig-full1}
\includegraphics[width=9cm,height=7cm]{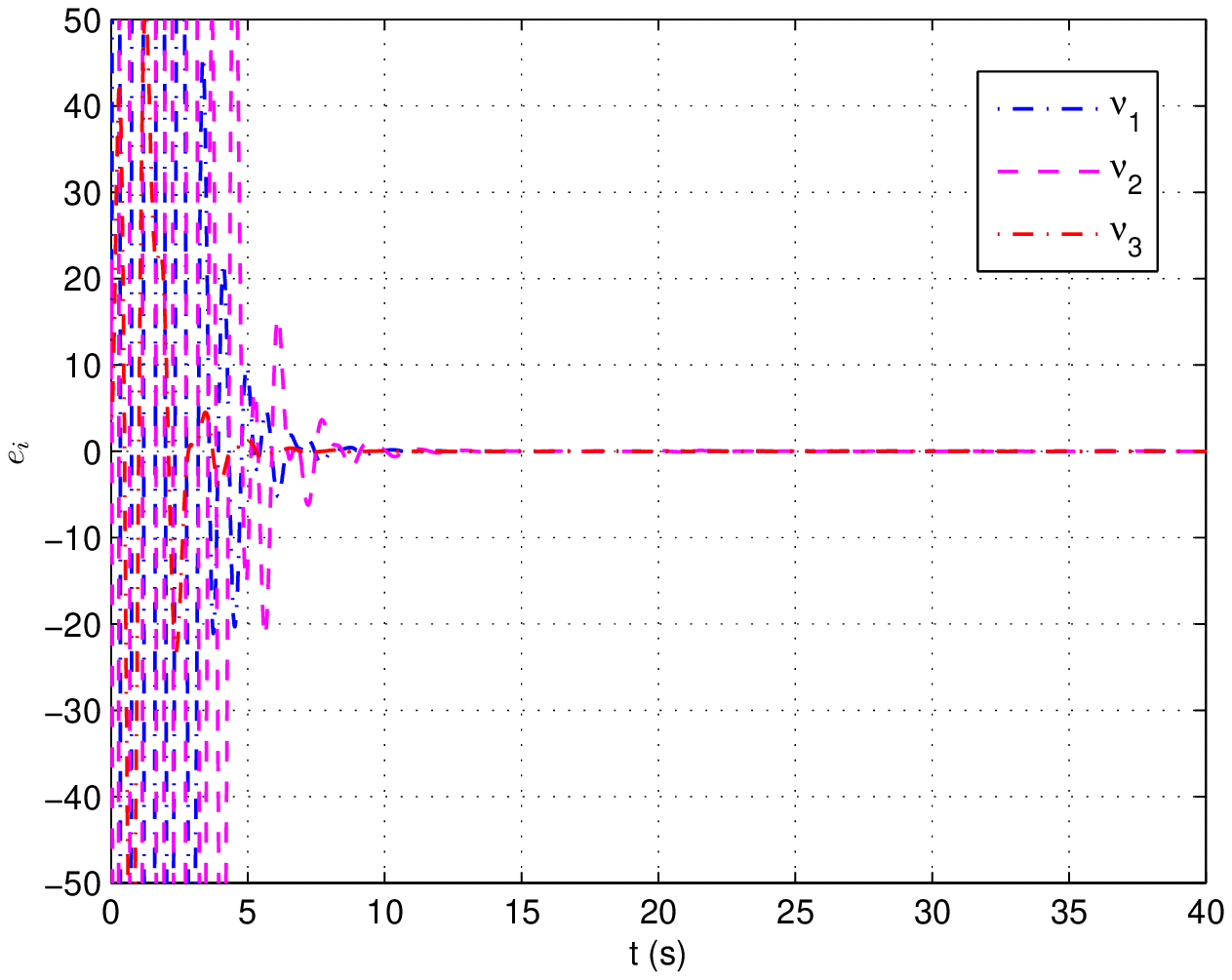}
  \caption{Error convergence of system \eqref{eq:state}, \eqref{eq:disturbance},
  and \eqref{eq:leader}
  under the observer-based controller \eqref{eq:control-full}
  for Theorem \ref{thm:output}}
\label{fig-full2}
 \end{center}
 \end{figure}

Figs. \ref{fig-full1} and \ref{fig-full2} show, respectively,
the state convergence and the error convergence of system
\eqref{eq:state}, \eqref{eq:disturbance}, and \eqref{eq:leader}
under the observer-based controller \eqref{eq:control-full}.
We see that coordinated output regulation is realized even when there exists
multiple heterogenous dynamics and the information interactions are switching.
This agrees with Theorem \ref{thm:output}.

\textbf{Example 2}

We next give an example to validate Corollary \ref{thm:special1}. In this section,
the dynamics of the agents are described as
$A_1=\left[\begin{array}{ccc}0 & 3 & 0\\
0 & 0 & 2\\ 0 & -1 & 0 \end{array}\right]$,
$B_1=\left[\begin{array}{c}0 \\ 0 \\ 1\end{array}\right]$,
$C_{s1}=C_{d1}=D_{s1}=\left[\begin{array}{ccc}1 & 1 & 1\end{array}\right]$,
$A_2=\left[\begin{array}{cc}0 & 1 \\
0 & 0 \end{array}\right]$, $B_2=\left[\begin{array}{c}0 \\ 1\end{array}\right]$,
$C_{s2}=C_{d2}=D_{s2}=\left[\begin{array}{ccc}1 & 0\end{array}\right]$,
$A_3=\left[\begin{array}{cc}0 & 1 \\
 -2& -2 \end{array}\right]$, $B_3=\left[\begin{array}{c}0 \\ 1\end{array}\right]$,
 $C_{s3}=C_{d3}=D_{s3}=\left[\begin{array}{ccc}1 & 0\end{array}\right]$.
The dynamics of the individual autonomous exosystem are described as
$\omega_1(t)=0$, $\omega_2(t)=0$, and $\omega_3(t)=0$.
The dynamics of the group autonomous exosystem are described as
$A_0=\left[\begin{array}{cc}0 & 1\\
-1 & 0  \end{array}\right]$,
 $C_0=\left[\begin{array}{cc}1 & 0 \end{array}\right]$, $D_{0}=-C_0$.
\begin{figure}
\begin{center}
\includegraphics[width=9cm,height=7cm]{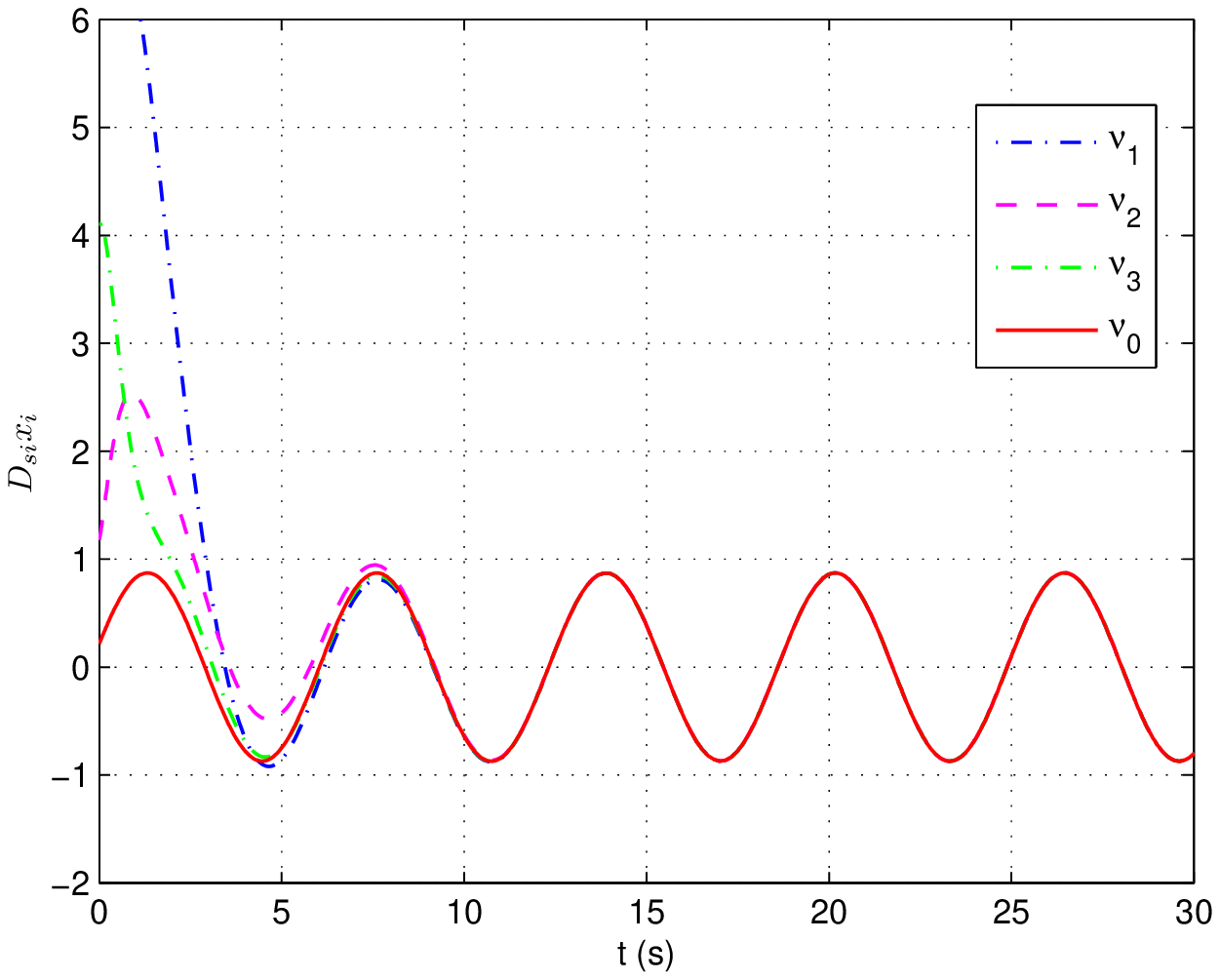}
  \caption{Output convergence of system \eqref{eq:state}, \eqref{eq:disturbance},
and \eqref{eq:leader} under the observer-based controller \eqref{eq:control-special1}
for Corollary \ref{thm:special1}}
\label{fig1}
\includegraphics[width=9cm,height=7cm]{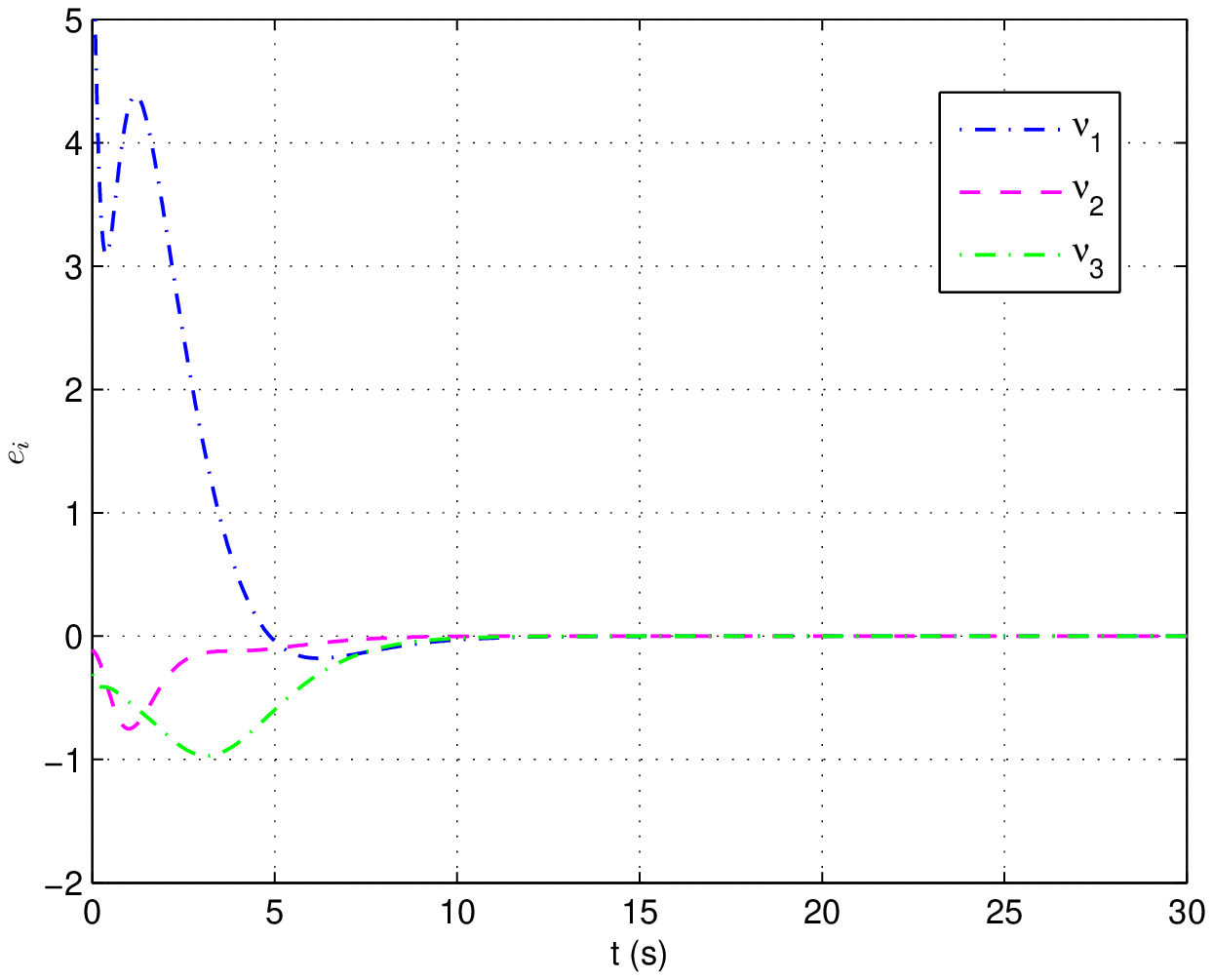}
  \caption{Error convergence of system \eqref{eq:state}, \eqref{eq:disturbance},
and \eqref{eq:leader} under the observer-based controller \eqref{eq:control-special1}
for Corollary \ref{thm:special1}}
\label{fig2}
 \end{center}
 \end{figure}

Following the design scheme proposed in Section \ref{sec:cor1}, for the solutions of regulator
equations \eqref{eq:regulator-cor1},
we have that $F_1=\left[\begin{array}{ccc} -1 & -4.5 & -6 \end{array}\right]$,
$\Pi_{21}=\left[\begin{array}{cc}1.0345 & -0.4138\\
0.1379 & 0.3448\\ -0.1724 & 0.0690  \end{array}\right]$,
 $\Gamma_{21}=\left[\begin{array}{cc}0.0690 & 0.1724 \end{array}\right]$ for agent $\nu_1$,
 $F_2=\left[\begin{array}{cc} -2 & -3 \end{array}\right]$,
 $\Pi_{22}=\left[\begin{array}{cc}1 & 0\\
0 & 1  \end{array}\right]$,
 $\Gamma_{22}=\left[\begin{array}{cc}-1 & 0 \end{array}\right]$ for agent $\nu_2$,
 $F_3=\left[\begin{array}{cc} 0 & -1 \end{array}\right]$,
 $\Pi_{23}=\left[\begin{array}{cc}1 & 0\\
0 & 1  \end{array}\right]$,
 $\Gamma_{23}=\left[\begin{array}{cc}1 & 2 \end{array}\right]$ for agent $\nu_3$.
We also have $K_{a1}=[-0.75,-4,-1.25]\T$, $K_{a2}=[-3,-2]\T$, $K_{a3}=[-1,2]\T$
for \eqref{eq:x-w-hat-special1}, $\varepsilon=0.2$ for \eqref{eq:observer-x0-special1} and
$\theta=0.1$ for \eqref{eq:P-special1}.

Figs. \ref{fig1} and \ref{fig2} show, respectively,
the state convergence and the error convergence of system
\eqref{eq:state}, \eqref{eq:disturbance},
and \eqref{eq:leader} under the observer-based controller \eqref{eq:control-special1}.
We see that coordinated output regulation is realized even when there exists multiple
heterogenous dynamics and the information interactions are switching.
This agrees with Corollary \ref{thm:special1}.

\textbf{Example 3}

We give an example to validate Corollary \ref{thm:special2}, the dynamics of the agents are described as
$A_1=\left[\begin{array}{ccc}0 & 3 & 0\\
0 & 0 & 2\\ 0 & -1 & 0 \end{array}\right]$, $B_1=\left[\begin{array}{c}0 \\ 0 \\ 1\end{array}\right]$, $C_{s1}=C_{d1}=D_{s1}=\left[\begin{array}{ccc}1 & 1 & 1\end{array}\right]$.
$A_2=\left[\begin{array}{cc}0 & 1 \\
0 & 0 \end{array}\right]$, $B_2=\left[\begin{array}{c}0 \\ 1\end{array}\right]$, $C_{s2}=C_{d2}=D_{s2}=\left[\begin{array}{ccc}1 & 0\end{array}\right]$.
$A_3=\left[\begin{array}{cc}0 & 1 \\
 -2& -2 \end{array}\right]$, $B_3=\left[\begin{array}{c}0 \\ 1\end{array}\right]$, $C_{s3}=C_{d3}=D_{s3}=\left[\begin{array}{ccc}1 & 0\end{array}\right]$. The dynamics of the individual autonomous exosystems are described as $S_i=0$, $C_{wi}=D_{wi}=-1$, $i=1,2,3$, and $\omega_1(0)=-2$, $\omega_2(0)=-4$, and $\omega_3(0)=-6$.
The dynamics of the group autonomous exosystem are described as $A_0=\left[\begin{array}{cc}0 & 1\\
-1 & 0  \end{array}\right]$,
 $C_0=\left[\begin{array}{cc}1 & 0 \end{array}\right]$, $D_{0}=-C_0$.

\begin{figure}
\begin{center}
\includegraphics[width=9cm,height=7cm]{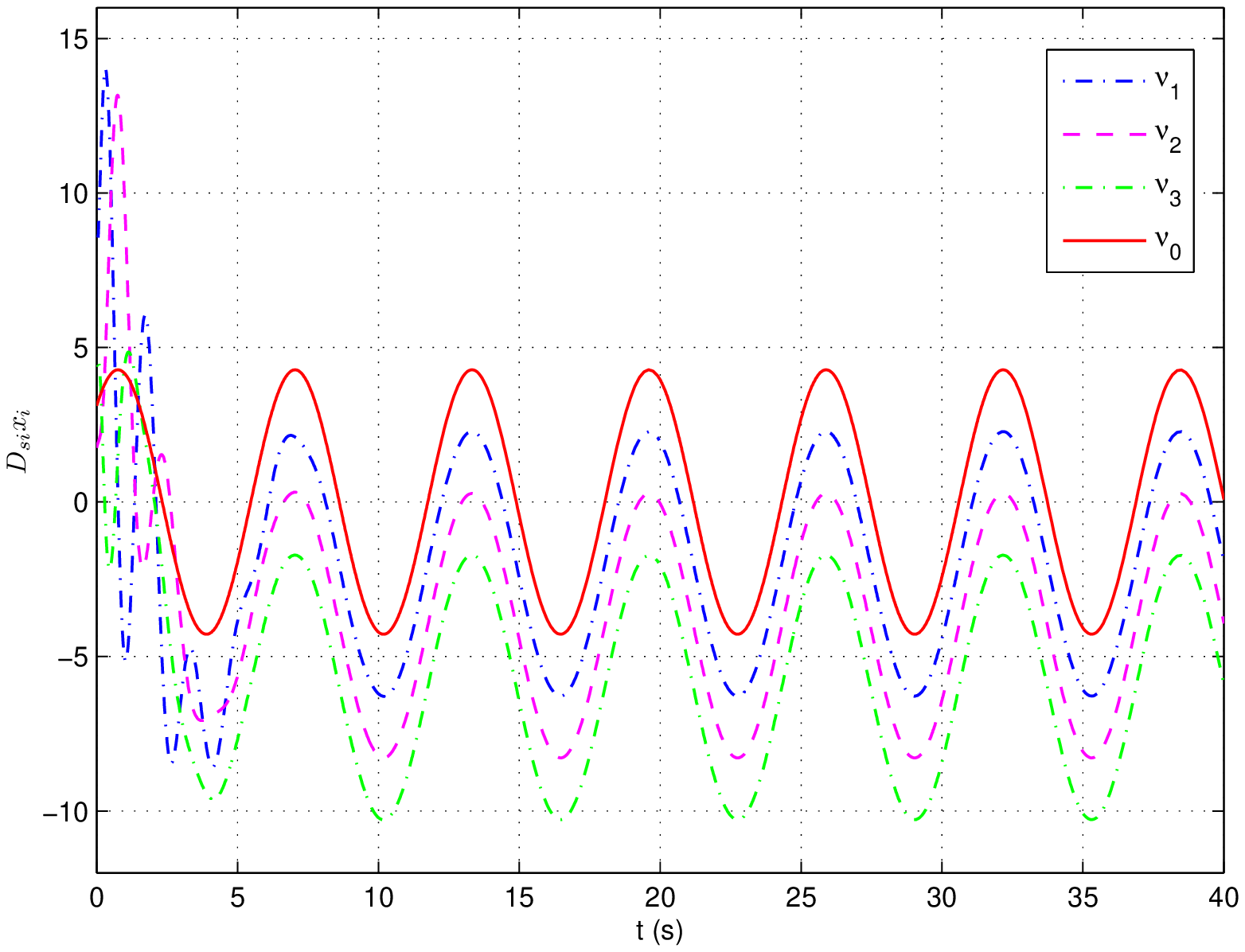}
  \caption{Output convergence of system \eqref{eq:state}, \eqref{eq:disturbance},
and \eqref{eq:leader} under the observer-based controller \eqref{eq:control-special2} for Corollary \ref{thm:special2}}
\label{fig3}
\includegraphics[width=9cm,height=7cm]{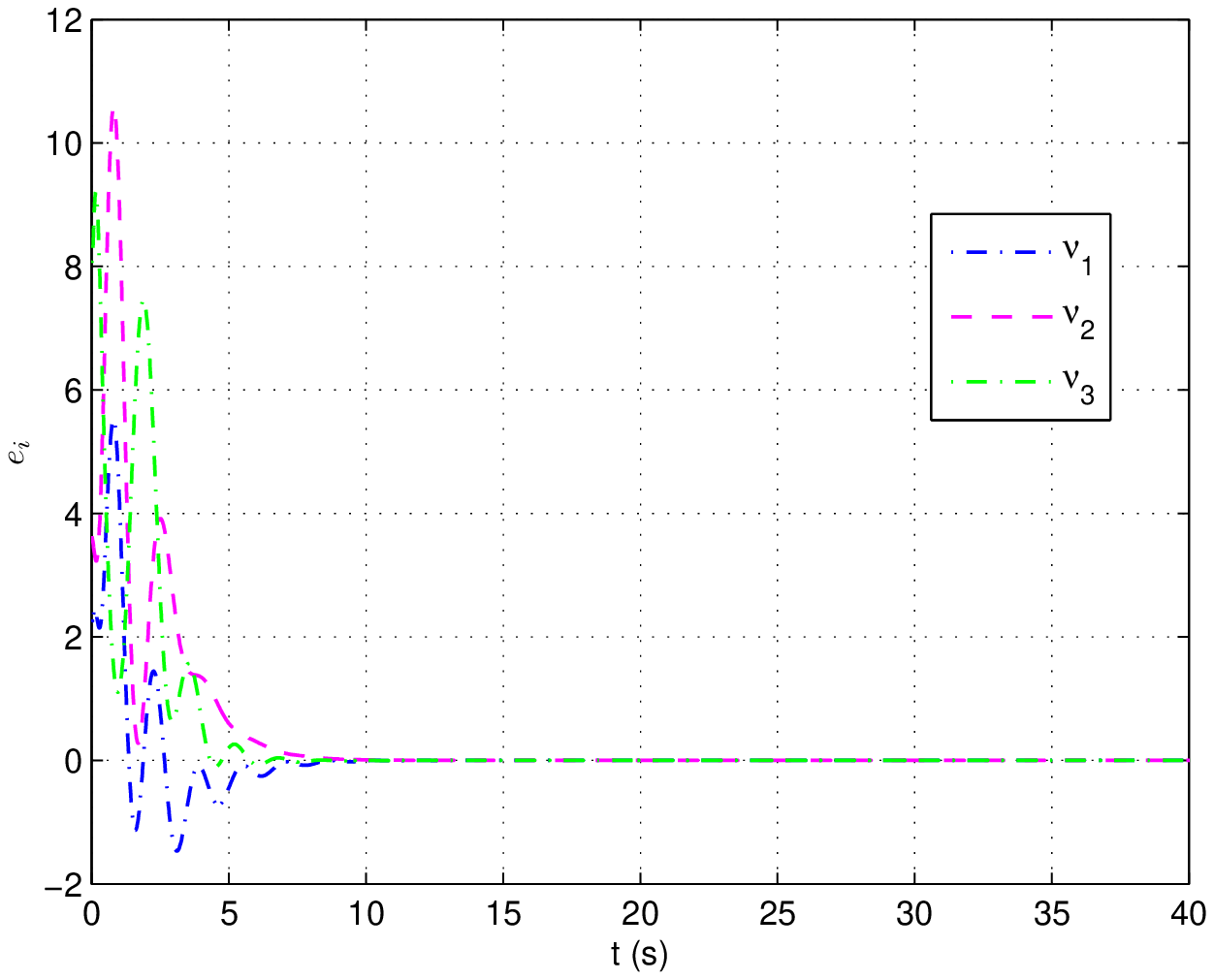}
  \caption{Error convergence of system \eqref{eq:state}, \eqref{eq:disturbance},
and \eqref{eq:leader} under the observer-based controller \eqref{eq:control-special2} for Corollary \ref{thm:special2}}
\label{fig4}
 \end{center}
 \end{figure}

Following the design scheme proposed in Section \ref{sec:cor2}, for the solutions of regulator
equations \eqref{eq:regulator-cor2}, we have that $F_1=\left[\begin{array}{ccc} -1 & -4.5 & -6 \end{array}\right]$,
$\Pi_{11}=\left[\begin{array}{c}1 \\
0 \\ 0  \end{array}\right]$,
 $\Gamma_{11}=0$, $\Pi_{21}=\left[\begin{array}{cc}1.0345 & -0.4138\\
0.1379 & 0.3448\\ -0.1724 & 0.0690  \end{array}\right]$,
 $\Gamma_{21}=\left[\begin{array}{cc}0.0690 & 0.1724 \end{array}\right]$ for agent $\nu_1$,
$F_2=\left[\begin{array}{cc} -2 & -3 \end{array}\right]$,
$\Pi_{12}=\left[\begin{array}{c}1 \\
0   \end{array}\right]$,
 $\Gamma_{12}=0$, $\Pi_{22}=\left[\begin{array}{cc}1 & 0\\
0 & 1  \end{array}\right]$,
 $\Gamma_{22}=\left[\begin{array}{cc}-1 & 0 \end{array}\right]$ for agent $\nu_2$,
 $F_3=\left[\begin{array}{cc} 0 & -1 \end{array}\right]$,
 $\Pi_{13}=\left[\begin{array}{c}1 \\
0   \end{array}\right]$,
 $\Gamma_{13}=-2$,
 $\Pi_{23}=\left[\begin{array}{cc}1 & 0\\
0 & 1  \end{array}\right]$,
 $\Gamma_{23}=\left[\begin{array}{cc}1 & 2 \end{array}\right]$ for agent $\nu_3$.
We also have $\varepsilon=0.2$ for \eqref{eq:observer-special2},
$\theta=0.1$ for \eqref{eq:P-special2}, and $K_{si}=1$, $i=1,2,3$
for \eqref{eq:w-hat-special2},

Figs. \ref{fig3} and \ref{fig4} show, respectively, the state convergence and the error convergence of system \eqref{eq:state}, \eqref{eq:disturbance},
and \eqref{eq:leader} under the observer-based controller \eqref{eq:control-special2}.
We see that coordinated output regulation is realized even when there exists multiple heterogenous dynamics and the information interactions are switching.
This agrees with Corollary \ref{thm:special2}.

\section{Conclusions}\label{sec:conclusion}
This paper studied the coordinated output regulation problem of multiple heterogeneous
linear systems. We first formulated the coordinated output regulation problem
and specified the information that is available for each agent.
 A high-gain based distributed observer and an individual observer were introduced
 for each agent and observer-based controllers were designed to solve the problem.
 The information interactions among the agents and the group autonomous exosystem
 were allowed to be switching over a finite set of fixed networks containing both
 the graph having a spanning tree and the graph having not. The relationship
 of the information interactions, the dwell time,
the non-identical dynamics of different agents,
and the high-gain parameters were also given. Simulations were given to
validate the theoretical results. Future directions include relaxing
the dwell-time assumption.

\bibliographystyle{elsart-num}
\bibliography{refs}

\end{document}